\documentclass[useAMS,usenatbib]{mn2e}

\usepackage{epsfig, bm}

\newcommand       \apj          {ApJ}
\newcommand       \apjl         {ApJL}
\newcommand       \aap          {A\&A}
\newcommand       \araa        {ARA\&A}

\newcommand       \mnras        {MNRAS}
\newcommand       \pasp       {PASP}

\def\s{{\rm\,s}}
\def\erg{{\rm\,erg}}
\def\cm{{\rm\,cm}}

\def\pc{{\rm\,pc}}
\def\kev{{\rm\,keV}}

\def\deg{{\rm \, deg}}
\def\sr{{\rm \, sr}}

\def\K{{\rm\,K}}
\def\yr{{\rm\,yr}}

\def\rt{R_{\rm T}}
\def\rs{R_{\rm S}}

\def\rmax{R_{\rm max}}
\def\mbh{M_{\rm BH}}
\def\msun{M_\odot}
\def\mstar{M_\star}
\def\rstar{R_\star}
\def\rp{R_{\rm p}}
\def\rin{R_{\rm in}}

\def\rlso{R_{\rm LSO}}
\def\rl{R_{\rm L}}
\def\vstar{v_\star}

\def\vp{v_{\rm p}}
\def\tfall{t_{\rm fallback}}

\def\trat{\left(\frac{t}{0.1\yr}\right)}
\def\hnurat{\left(\frac{\langle h\nu\rangle}{0.1\kev}\right)}
\def\mdotrat{\left(\frac{\dot{M}_{\rm fallback}}{\dot{M}_{\rm Edd}}\right)}
\def\mbhrat{M_6}
\def\rprat{R_{\rm p, 3\rs}}
\def\mstarrat{m_\star}
\def\rstarrat{r_\star}
\def\lesssim{\mathrel{\hbox{\rlap{\hbox{\lower3.5pt\hbox{$\sim$}}}\hbox{$<$}}}}
\def\gtrsim{\mathrel{\hbox{\rlap{\hbox{\lower3.5pt\hbox{$\sim$}}}\hbox{$>$}}}}

\title[Optical Signatures of Tidally Disrupted Stars]{Optical Flares from the Tidal Disruption of Stars by Massive Black Holes}
\author[L. E. Strubbe and E. Quataert]{
Linda E. Strubbe$^{1}$\thanks{E-mail:linda@astro.berkeley.edu},
Eliot Quataert$^{1}$ \\ 
$^{1}$Astronomy Department and Theoretical Astrophysics Center, 
601 Campbell Hall, University of California, Berkeley CA, 94720, USA\\}

\date{Accepted for publication in MNRAS}

\begin{document}

\pagerange{\pageref{firstpage}--\pageref{lastpage}} \pubyear{2009}
\maketitle

\label{firstpage}

\begin{abstract}
A star that wanders too close to a massive black hole (BH) is
  shredded by the BH's tidal gravity.  Stellar gas falls back to the
  BH at a rate initially exceeding the Eddington rate, releasing a
  flare of energy.  
In anticipation of upcoming transient surveys, 
we predict the light curves and spectra of tidal flares as a function
of time, highlighting the unique signatures of tidal flares at optical
and near-infrared wavelengths. A reasonable fraction of the gas
initially bound to the BH is likely blown away when the
fallback rate is super-Eddington at early times.  This outflow
produces an optical luminosity comparable to that of a supernova; such
events have durations of $\sim 10$ days and may have been missed in
supernova searches that exclude the nuclear regions of galaxies.  When
the fallback rate subsides below Eddington, the gas accretes onto the
BH via a thin disk whose emission peaks in the UV to soft X-rays.
Some of this emission is reprocessed by the unbound stellar debris,
producing a spectrum of very broad emission lines (with no
corresponding narrow forbidden lines). These lines are the strongest
for BHs with $\mbh \sim 10^5 - 10^6 \msun$ and thus optical surveys
are particularly sensitive to the lowest mass BHs in galactic nuclei.
Calibrating our models to ROSAT and GALEX observations, we predict
detection rates for Pan-STARRS, PTF, and LSST
and highlight some of the observational challenges associated with
studying tidal disruption events in the optical.  Upcoming surveys
such as Pan-STARRS should detect at least several events per year, and
may detect
many more if current models of outflows during super-Eddington
accretion are reasonably accurate.
These surveys
will significantly improve our knowledge of stellar dynamics in
galactic nuclei, the physics of super-Eddington accretion, the
demography of intermediate mass BHs, and the role of tidal disruption
in the growth of massive BHs.
\end{abstract}


\section{Introduction}\label{intro}

Stellar orbits in the center of a galaxy are not static, and sometimes
stars walk into trouble.  If an unlucky star passes within $\rt \sim
\rstar(\mbh/\mstar)^{1/3}$ of the galaxy's central black hole (BH), the
BH's tidal gravity exceeds the star's self-gravity, and the star is
disrupted.  For BHs with $\mbh \le 10^8M_\odot$, the disruption of a
solar-type star occurs outside the horizon and is likely accompanied
by a week- to year-long electromagnetic flare (e.g., \citealt{rees}).

Gravitational interactions between stars ensure that all supermassive
BHs tidally disrupt nearby stars (e.g., \citealt{magtrem}).  The
scattering process might be accelerated by resonant relaxation very
close to the BH \citep{rauchtrem}, or interactions with
``massive perturbers'' like a massive accretion disk \citep{zhao} or giant
molecular clouds \citep{perets}.  In addition, the galactic potential
may be triaxial so stars need not be scattered at all: they may simply
follow their chaotic orbits down to $\sim\rt$ \citep{mp}.  Given
these uncertainties, predictions for the timescale between tidal
disruptions in a given galaxy range from $10^3$ to $10^6$ years.  The
rate remains uncertain, but tidal disruption must occur.

Indeed, a handful of candidate events have been detected to date.  The
accreting stellar debris is expected to emit blackbody radiation from
very close to the BH, so X-ray and UV observations probe the
bulk of the emission.  Several candidate tidal disruption events were
discovered in the ROSAT All-Sky Survey
\citep[see][]{komossa} and the XMM-Newton Slew Survey \citep{esquej}; 
the GALEX Deep Imaging Survey has so far yielded three candidates
\citep{gezari06,gezari08,gezari09}.
For ROSAT, these detections are consistent with a rate $\sim
10^{-5}\yr^{-1}$ per galaxy \citep{donley}, but the data are sparse.
However, we are entering a new era of transient surveys: in the
optical, surveys like Pan-STARRS (PS1, then all four telescopes)
\citep[e.g.,][]{magnier}, the Palomar Transient Factory \citep{rau},
and later the Large Synoptic Survey Telescope
will have fast cadence, wide fields of view, and unprecedented
sensitivity.  Wide-field transient surveys with rapid cadence are also
planned at other wavelengths, including the radio (e.g., LOFAR and the
ATA), near-infrared (e.g., SASIR), and hard X-rays (e.g., EXIST).  How many
tidal flares these surveys find depends on their luminosity and
spectra as a function of time.

In this paper, we predict the light curves and spectra of tidal
disruption events as a function of time.  Since the early work on
tidal disruption \citep[e.g.,][]{rees}, it has been well-appreciated that
the bulk of the emission occurs in the UV and soft X-rays, with a
possible extension to harder X-rays.\footnote{Such a hard X-ray component
may be detectable with upcoming all-sky X-ray surveys like the proposed EXIST mission
\citep{grindlay}.
However, we choose not to include predictions for hard X-rays in our calculations:
one could draw analogy to the hard X-ray power-law tail observed in AGN spectra, but since the origin
of this feature is uncertain,
it is difficult to make firm theoretical predictions for tidal disruption events.}
Taking into account only this
emission, optical wavelengths are not the most promising for detecting
tidal flares, because the blackbody temperature of the inner accretion
disk is $\sim 3\times 10^5\K$.  We show, however, that there are two
additional sources of optical emission that likely dominate the
optical flux in many, though not all, cases: (1) emission produced by
a super-Eddington outflow at early times and (2) emission produced by
the irradiation and photoionization of unbound stellar debris
\citep[see also the earlier work of][]{bogdanovic}.  In the latter
case, much of the optical emission is in the form of very broad
emission lines, while in the former it is primarily continuum
(although some lines may also be present).  Throughout this paper, we
typically discuss these two sources of emission separately, largely
because the physics of the photoionized stellar debris is more secure
than that of the super-Eddington outflows.

The remainder of this paper is organized as follows.  In
\S\S\ref{outflows}, \ref{diskmodel} and \ref{unbound}, we describe our
models for the polar super-Eddington outflow, accretion disk and the
equatorial unbound material, respectively; then in
\S\ref{emissionprops} we calculate the luminosity and spectral
signatures of tidal disruption events.  We predict detection rates in
\S\ref{rates}, and summarize and discuss our results in
\S\ref{discussion}.  \S\S \ref{data} and \ref{discussion} include a
discussion of our models in the context of ROSAT and GALEX
observations of tidal flare candidates.

\section{The Initially Bound Material} \label{bound}

We consider a star approaching the BH on a parabolic orbit with
pericenter distance $\rp\le\rt$.  Once the star reaches the vicinity
of the tidal radius ($\rt$), the tidal gravity stretches it radially
and compresses it vertically and azimuthally.  The acceleration is $a
\sim (G\mbh/\rp^2)(\rstar/\rp)$ and acts for a dynamical time $t_{\rm
  p} \sim (G\mbh/\rp^3)^{-1/2}$ near pericenter, resulting in velocity
perturbations $\Delta v_{\rm p} \sim a t_{\rm p} \sim
\vstar(\rt/\rp)^{3/2}$, where $\rstar$ is the star's radius and
$\vstar$ is the star's escape velocity.  The change in velocity
$\Delta v_{\rm p}$ is smaller than the star's orbital velocity at
pericenter, $v_{\rm p} \equiv (2G\mbh/\rp)^{1/2}$, by a factor of
$\rstar/\rp$.

Because $\Delta v_{\rm p}$ is at least as large as the sound speed
inside the star, the stellar gas may shock vertically and azimuthally
\citep[e.g.,][]{brassart, guillochon}.
Once the shredded star passes through pericenter, the compression
subsides and the star re-expands, cooling adiabatically; thermal
pressure becomes negligible and the particles travel away from the
BH ballistically.  We assume that the particle trajectories
become ballistic when the star passes through pericenter.  At that
time, the particles have perturbed azimuthal, vertical, and radial velocities
$\sim\Delta \vp$.

The particles have a range in specific energy $\mathcal{E} \sim
\pm 3(G\mbh/\rp)(\rstar/\rp)$ \citep[e.g.,][]{lacy,li},  
due to their relative locations in the BH's potential well and
differences in their azimuthal speeds.  Initially, approximately half of the
stellar mass is bound and half is unbound \citep
{lacy, evkoch}.  After a time
\begin{eqnarray}\label{tfallback}
t_{\rm fallback} & \sim &
\frac{2\pi}{6^{3/2}}\left(\frac{\rp}{\rstar}\right)^{3/2} t_{\rm p} \\
& \sim & 20 \mbhrat^{5/2}\rprat^3 \rstarrat^{-3/2} {\, \rm min} \, ,
\nonumber
\end{eqnarray}
the most bound material returns to pericenter.  (Here we have defined
$\mbhrat \equiv \mbh/10^6M_\odot$, $\rprat \equiv \rp/3\rs$, and $\rstarrat \equiv \rstar/R_\odot$.)
Less bound gas follows, at a rate
\begin{equation}
\dot{M}_{\rm fallback} \approx \frac{1}{3}\frac{\mstar}{t_{\rm fallback}}\left(\frac{t}{t_{\rm fallback}}\right)^{-5/3}
\label{mdotfallback}
\end{equation}
\citep{rees,phinney}.  There will be some deviations from this
canonical $t^{-5/3}$ scaling at early times, depending on the precise
structure of the star \citep{lodato,ramirez}, but we use equation
(\ref{mdotfallback}) for simplicity.  As matter returns to pericenter,
it shocks on itself, converting most of its bulk orbital energy to
thermal energy \citep[see][]{kochanek}.  The viscous time is typically shorter than the
fallback time, so at least some of the matter begins to accrete.

For $\mbh \lesssim {\rm few} \times 10^7 M_\odot$, the mass fallback
rate predicted by equations (\ref{tfallback}) \& (\ref{mdotfallback})
can be much greater than the Eddington rate $\dot{M}_{\rm Edd}$ for a
period of weeks to years; here $\dot{M}_{\rm Edd} \equiv 10L_{\rm  Edd}/c^2$, $L_{\rm Edd}$
is the Eddington luminosity, and 0.1 is the assumed efficiency of converting accretion
power to luminosity.
The fallback rate only falls below the Eddington rate at 
a time
\begin{equation}
t_{\rm Edd} \sim 0.1 \, \mbhrat^{2/5}\rprat^{6/5}\mstarrat^{3/5}\rstarrat^{-3/5}\yr,
\label{tEdd}
\end{equation}
where $m_\star \equiv M_\star/M_\odot$.  While the fallback rate is
super-Eddington, the stellar gas returning to pericenter is so dense
that it cannot radiate and cool.  In particular, the time for photons
to diffuse
out of the gas is longer than both the inflow time in the disk
and the dynamical time characteristic of an outflow.  The gas is
likely to form an advective accretion disk accompanied by powerful
outflows \citep[e.g.,][]{ohsuga}, although the relative importance of
accretion and outflows in this phase is somewhat uncertain (see
\S\ref{discussion}). Later, when $\dot{M}_{\rm fallback} <
\dot{M}_{\rm Edd}$ ($t > t_{\rm Edd}$), the outflows subside, and the
accretion disk can radiatively cool and becomes thin.  

In \S\ref{outflows}, we describe our model for the super-Eddington
outflows, and in \S\ref{diskmodel}, we describe our model for the
accretion disk.  We discuss uncertainties in these models in
\S\ref{discussion}.
 
 \subsection{Super-Eddington Outflows}\label{outflows}

 When the fallback rate to pericenter is super-Eddington, radiation
 produced by the shock and by viscous stresses in the rotating disk is
 trapped by electron scattering.  By energy conservation, this
 material is initially all bound to the BH, but it is only weakly bound
 because the radiation cannot escape and because the material originated on
 highly eccentric orbits.  Some fraction of the returning gas is thus
 likely unbound \citep[see, e.g., the simulations of][]{ayal}, with
 energy being conserved as other gas accretes inward
 \citep{blandfordbegelman}.  If the outflow's covering fraction is
 high, most of the radiated power will be emitted from the outflow's
 photosphere, which can be far outside $\sim \rp$ \citep{loeb}.  We
 now estimate the properties of this outflowing gas (see
 \citealt{rossi} for related estimates in the context of
 short-duration gamma-ray bursts).

 In our simplified scenario, stellar debris falls back at close to the
 escape velocity and shocks at the launch radius $\rl \sim 2\rp$,
 converting bulk kinetic energy to radiation:
\begin{equation}
aT_{\rm L}^4 \sim \frac{1}{2}\rho_{\rm fallback, L} v_{\rm esc, L}^2 \, ,
\end{equation}
where $T_{\rm L}$ is the temperature at $\rl$,
$\rho_{\rm fallback, L} \sim \dot{M}_{\rm fallback}/(4\pi\rl^2v_{\rm esc, L})$
is the density of gas at $\rl$, and $v_{\rm esc, L} \equiv (2G\mbh/\rl)^{1/2}$.
 Outflowing gas is launched from $\rl$ at a rate
\begin{equation}
 \dot{M}_{\rm out} \equiv f_{\rm out}\dot{M}_{\rm fallback} 
\label{mdotout}
\end{equation}
and with
 terminal velocity 
\begin{equation}
  v_{\rm wind} \equiv f_v v_{\rm esc}(\rl).
\label{vout}
\end{equation}
We approximate the outflow's geometry as spherical, with a density
profile
\begin{equation}
\rho(r) \sim \frac{\dot{M}_{\rm out}}{4\pi r^2 v_{\rm wind}} 
\end{equation}
inside the outflow where $r \lesssim R_{\rm edge} \equiv v_{\rm
  wind}t$; the density falls quickly to zero at $\sim R_{\rm edge}$.
We define the trapping radius
$R_{\rm trap}$ via $R_{\rm  trap}\rho(R_{\rm trap})\kappa_{\rm s} \sim c/v_{\rm wind}$ (where
$\kappa_{\rm s}$ is the opacity due to electron scattering):
inside $R_{\rm trap}$, the gas is too optically thick for photons to escape
and so the outflowing gas expands adiabatically.  Because the outflow remains
supported by radiation pressure, $T \propto \rho^{1/3}$.
The photosphere of the outflow $R_{\rm ph}$ is where $R_{\rm ph}\rho(R_{\rm ph})\kappa_{\rm s} \sim 1$.
Because $v_{\rm wind}$ is
likely not much smaller than $c$, $R_{\rm trap} \sim R_{\rm ph}$; we
thus neglect any deviations from adiabaticity between $R_{\rm trap}$
and $R_{\rm ph}$.

At the earliest times for small $\mbh$ and $\rp$, the fallback rate
can be so large and the density so high that the edge of the outflow
limits the location of the photosphere to be just inside $R_{\rm edge}$.
In that case, the density of the photosphere is still given
by $\rho(R_{\rm ph}) \sim (\kappa_{\rm s}R_{\rm edge})^{-1}$; lacking
a detailed model, we assume that the photospheric gas near the edge is
on the same adiabat as the rest of the gas, so that
\begin{eqnarray}
T_{\rm ph} & \sim & T_{\rm L}\left[\frac{\rho(R_{\rm ph})}{\rho_{\rm fallback,L}}\right]^{1/3} \\
& \sim & 3\times10^4 f_{v}^{-1/3} \mbhrat^{1/36}\rprat^{-1/8}\mstarrat^{-1/12}\rstarrat^{1/12}\left(\frac{t}{\rm day}\right)^{-7/36} \K \, . \nonumber
\label{tphedge}
\end{eqnarray}
Note that the photospheric temperature during the edge-dominated phase
is essentially independent of all parameters of the disruption (e.g.,
$M_{\rm BH}$, $\rp$, etc.), and is only a weak function of time. The
total luminosity during this phase grows as $L \propto t^{11/9}$ while
the luminosity on the Rayleigh-Jeans tail increases even more rapidly,
$\nu L_\nu \propto t^{65/36}$.
After a time
\begin{equation}\label{tedge}
t_{\rm edge} \sim 1 \,\,
f_{\rm out}^{3/8}f_v^{-3/4}\mbhrat^{5/8}\rprat^{9/8}\mstarrat^{3/8}\rstarrat^{-3/8} \,{\rm day} \, ,
\end{equation}
the density falls sufficiently that the photosphere lies well inside $R_{\rm edge}$; the photosphere's radius is then
\begin{equation}
R_{\rm ph} \sim 4 f_{\rm out}f_v^{-1}\mdotrat \rprat^{1/2}\rs 
\label{rph}
\end{equation}
and its temperature is
\begin{equation}\label{tphnoedge}
T_{\rm ph} \sim  2\times10^5 f_{\rm out}^{-1/3}f_v^{1/3} \mdotrat^{-5/12}\mbhrat^{-1/4}\rprat^{-7/24} \,\, \K \, .
\end{equation}
The adiabatically expanding outflow preserves the photon distribution
function generated in the shock and accretion disk close to the
BH. Estimates indicate that this gas is likely to be close to thermal
equilibrium and thus we assume that the escaping photons have a
blackbody spectrum
\begin{equation}\label{Loutflow}
\nu L_\nu \sim 4\pi^2 R_{\rm ph}^2 \nu B_\nu(T_{\rm ph}) \, .
\end{equation}
When the photosphere lies inside the edge of the outflow (i.e., $t > t_{\rm edge}$ so $R_{\rm ph} < v_{\rm wind} t$),
equations (\ref{rph}) and (\ref{tphnoedge}) imply that the total luminosity of the outflow is
\begin{equation}
L \sim 10^{44} f_{\rm out}^{2/3}f_{v}^{-2/3} \mbhrat^{11/9} \rprat^{1/2} \mstarrat^{1/3}\rstarrat^{-1/3}
\left(\frac{t}{\rm day}\right)^{-5/9} \erg\s^{-1}.
\label{lph}
\end{equation}
The total luminosity of the outflow is thus of order the Eddington
luminosity: see Figure \ref{Lpeakfig}, discussed in
\S\ref{outflow_emission}.  Note that the total luminosity decreases
for lower outflow rates, $L \propto f_{\rm out}^{2/3}$, because the
photosphere's surface area is smaller.  The luminosity on the
Rayleigh-Jeans tail (generally appropriate for optical and
near-infrared wavelengths) declines even faster for lower $\dot M_{\rm
  out}$, scaling as
\begin{equation}
\label{lphRJ}
\nu L_\nu \propto f_{\rm out}^{5/3} f_v^{-5/3} \, .
\end{equation}
These relations only apply if $R_{\rm ph} \gtrsim R_{\rm L}$ because
otherwise the outflow is optically thin; we impose this lower limit to
$R_{\rm ph}$ in our numerical solutions described later.

\subsection{The Accretion Disk}\label{diskmodel}
  
We now consider the bound stellar debris that accretes onto the BH.
After shocking at pericenter, this gas circularizes and viscously
drifts inward, forming an accretion disk.  The disk extends from $\sim
2 \rp$ down to the last stable orbit, $R_{\rm LSO}$. We expect 
the viscous time in the disk to be substantially
shorter than the fallback time for at least a few years
(see \citealt{ulmer99} for the case of $\rp =\rt$, assuming a thick disk),
and check this expectation at the end of \S\ref{diskmodel}; we
thus assume that accretion during this period proceeds at
$\dot{M} \simeq (1-f_{\rm  out})\dot{M}_{\rm fallback}$.  
During the super-Eddington phase, the
time for photons to diffuse out of the disk is longer than the viscous
time, and so the disk is thick and advective.  In contrast, at later
times when $\dot{M}_{\rm fallback} \lesssim \dot{M}_{\rm Edd}$, the
disk is thin and can cool by radiative diffusion.  We derive an
analytic ``slim disk'' model \citep[similar to the numerical work
of][]{abram} to describe the structure of the disk in both regimes.

To calculate the disk's properties, we solve the equations of
conservation of mass, momentum, and energy: 
\begin{eqnarray}
\dot{M} = -4\pi R H \rho v_r \, , \\
v_r = -\frac{3}{2}\frac{\nu}{R}\frac{1}{f} \, , \\
q^+  =  q^- -\rho T v_r \frac{s}{R} \, , \label{entropy}
\end{eqnarray}
where we have approximated the radial entropy gradient as
$\partial{s}/\partial{R} \sim -s/R$.  Here $\dot{M}$ is the accretion
rate, $R$ is the cylindrical distance from the BH, $H$ is the disk
scale height, $\rho$ is the density, $v_r$ is the radial velocity, and
$T$ is the midplane temperature.  The no-torque boundary condition at
the inner edge of the disk implies $f \equiv 1-(R_{\rm LSO}/R)^{1/2}$.
We neglect gas pressure, since radiation pressure is dominant
throughout the disk for at least a few years; we further assume that
the viscous stress is proportional to the radiation pressure, so that
\citep{ss} $\nu = \alpha c_{\rm s} H$ with sound speed $c_s = (aT^4 /
3\rho)^{1/2}$ and $H=c_{\rm s}/\Omega_{\rm K}$, where $\Omega_{\rm K}
\equiv (G\mbh/R^3)^{1/2}$.  Simulations indicate that this assumption
is reasonable and that such disks are thermally stable \citep{hirose}.
The vertically integrated heating and cooling rates are given by
$2Hq^+ = 3G\mbh \dot{M}f/4\pi R^3$ and $2Hq^- = 8\sigma T^4/3\tau$,
where the half-height optical depth is $\tau = H\rho \kappa_{\rm s}$
and $\kappa_{\rm s}$ is the electron scattering opacity.  These
relations form a quadratic equation for the dimensionless quantity
$\kappa_{\rm s} aT^4/c \Omega_{\rm K} = 3 \tau(c_{\rm s}/c)$,
\begin{equation}
0=\left(\frac{\kappa_{\rm s} a T^4}{c\Omega_{\rm K}}\right)^2 
- \frac{4}{3\alpha} \left(\frac{\kappa_{\rm s} aT^4}{c\Omega_{\rm K}} \right)
- \frac{8f}{3\alpha^2}\left(\frac{10\dot{M}}{\dot{M}_{\rm Edd}}\right)^2
\left(\frac{R}{R_{\rm S}}\right)^{-2} \, .
\label{quadratic}
\end{equation}
Solving equation (\ref{quadratic}) yields
the effective temperature of the disk,
\begin{displaymath}
\sigma T_{\rm eff}^4 =  \frac{4\sigma T^4}{3\tau} =
\frac{3G\mbh\dot{M}f}{8\pi R^3} \times
\end{displaymath}
\begin{equation}
 \left[\frac{1}{2}+
\left\{\frac{1}{4}+\frac{3}{2}f
\left(\frac{10\dot{M}}{\dot{M}_{\rm Edd}}\right)^2\
\left(\frac{R}{\rs}\right)^{-2}\right\}^{1/2}\right]^{-1}.
\label{Tdisk}
\end{equation}
Combining this relation with equation (\ref{mdotfallback}), we
calculate the luminosity and spectrum of the disk as a function of
time, modeling it as a multicolor blackbody.

The solution to (\ref{quadratic}) also yields the disk scale height ratio,
\begin{displaymath}
\frac{H}{R} = \frac{3}{4}f\left(\frac{10\dot{M}}{\dot{M}_{\rm Edd}}\right)\left(\frac{R}{\rs}\right)^{-1} \times
\end{displaymath}
\begin{equation}
 \left[\frac{1}{2}+
\left\{\frac{1}{4}+\frac{3}{2}f
\left(\frac{10\dot{M}}{\dot{M}_{\rm Edd}}\right)^2\
\left(\frac{R}{\rs}\right)^{-2}\right\}^{1/2}\right]^{-1} \, .
\label{HoverR}
\end{equation}
The scale height $H \sim R$ while $\dot{M}_{\rm fallback} \gtrsim \dot{M}_{\rm Edd}$, and decreases as $t^{-5/3}$ at fixed $R$ at later times.
The viscous time at a radius $R$ in the disk is
\begin{equation}
t_{\rm visc} \sim \alpha^{-1}\left(\frac{G\mbh}{R^{3}}\right)^{-1/2} \left(\frac{H}{R}\right)^{-2} \, ,
\label{tvisc}
\end{equation}
which is $\sim \alpha^{-1}$ times the local dynamical time during the super-Eddington phase, and later increases as $t^{10/3}$ at fixed $R$.
For $\alpha \sim 0.1$, the viscous timescale evaluated at the disk's outer edge is shorter than the time $t$ since disruption 
for $\sim1-3\yr$; our assumption of steady-state accretion during this period is thus consistent.

\section{The Equatorial Unbound Material} \label{unbound}

While half of the initial star becomes bound to the BH during the
disruption, the other half gains energy and escapes from the BH on
hyperbolic trajectories.  From the viewpoint of the BH, this unbound
material subtends a solid angle $\Delta \Omega$, with a dispersion
$\Delta \phi$ in azimuth and a dispersion $\Delta i$ in orbital
inclination.  This material absorbs and re-radiates a fraction of the
luminosity from the 
accretion disk.\footnote{The polar outflow could also irradiate the
  unbound material, but it will have less of an effect because its
  spectrum is softer and its luminosity declines more rapidly.}  We
now estimate the dimensions of the unbound wedge.

\begin{figure}
\centerline{\epsfig{file=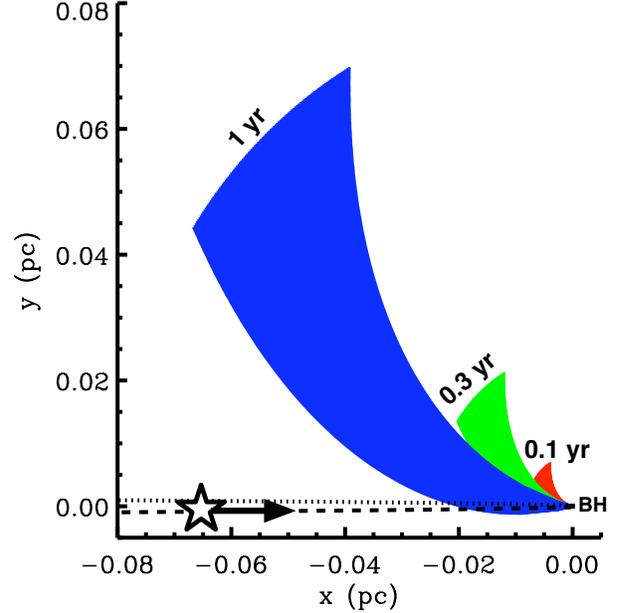, height=9cm}}
\vspace{-0.4cm}
\caption{Spatial diagram of the equatorial stellar debris in the plane defined by the original orbit of the star.  The unbound material is shown 0.1 yr, 0.3 yr, and 1 yr after the tidal disruption of a solar-type star by a $10^6 \msun$ BH at $\rp = 3\rs$.  The dashed and dotted curves indicate the incoming parabolic trajectory of the star and its continuation if the star were not disrupted.  The debris also has an inclination dispersion perpendicular to this plane of $\Delta i \sim 2\rstar/\rp$.
  \label{UBdiagram}}
\end{figure}

In the orbital plane at a fixed time $t \gtrsim t_{\rm fallback}$, the
unbound stellar debris lies along an arc, as the spread in specific
energy produces a spread in radius and azimuthal angle (see Fig. \ref{UBdiagram}).  The most
energetic particles escape on a hyperbolic orbit with eccentricity
$e_{\rm max}
\sim 1+6\rstar/\rp$.  These particles race away from the BH
at a substantial fraction of the speed of light,
\begin{equation}\label{vmax}
\frac{v_{\rm max}}{c} \sim \left(\frac{3\rstar}{\rp}\right)^{1/2}\frac{\vp}{c}
\sim 0.3 \mbhrat^{-1/2} \rprat^{-1}
\end{equation}
(ignoring relativistic effects) and lie furthest from the BH at a
distance 
\begin{equation} \label{rmax}
R_{\rm max} \sim 0.01 \mbhrat^{-1/2}\rprat^{-1}\rstarrat^{1/2}\trat \pc \, .
\end{equation}
They also have the smallest angle away from stellar
pericenter, $\phi_{\rm min} \sim f_\infty$, where $f_\infty$ obeys
$\cos f_\infty = -1/e_{\rm max}$ so that $\phi_{\rm min} \sim
\pi-(12\rstar/\rp)^{1/2}$ \citep[see also][]{khokhlov}. Particles with lower
energies and thus smaller eccentricities are closer to the BH and make
a larger angle relative to pericenter, up to $\phi \sim \pi$.  This
produces 
an azimuthal dispersion $\Delta \phi
\sim (12\rstar/\rp)^{1/2}$.

\begin{figure*}
\begin{center}
\epsfig{file=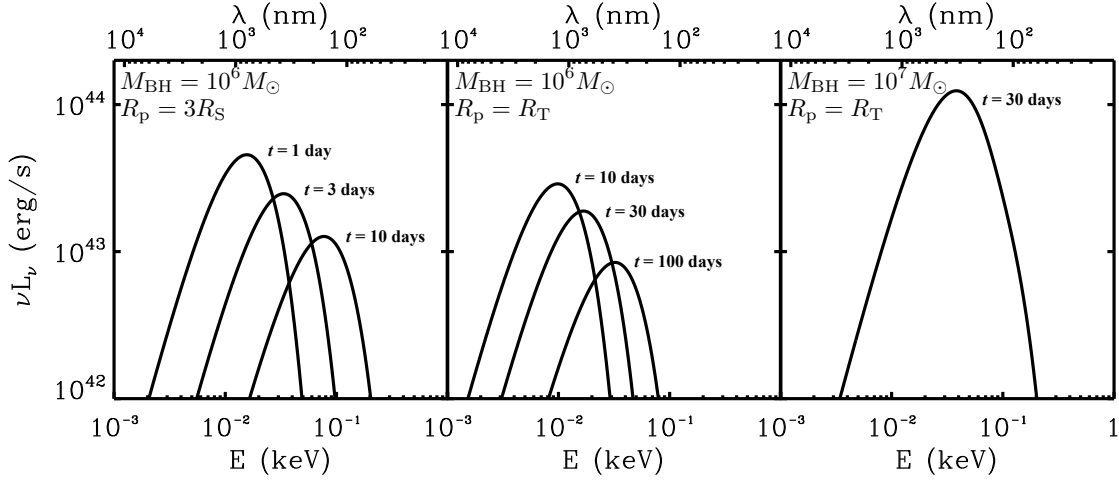, width=16cm}
\vspace{-0.4cm}
\end{center}
\caption{Spectral energy distributions for our three fiducial tidal
  disruption flares at several different times after disruption.
  These spectra include only the emission from the super-Eddington
  outflows (for $f_{\rm out} = 0.1$ and $f_v = 1$; see eqns
  [\ref{mdotout}] \& [\ref{vout}]), which dominate the emission at
  early times (see Fig. \ref{outflow_lc}).  For $M_{\rm BH} = 10^7 M_\odot$ and $\rp = \rt$ (right panel), the disk dominates the emission for $t \gtrsim 50$ days (Fig. \ref{outflow_lc}); this is why we do not plot the outflow emission at later times.
  \label{outflow_spect}}
\end{figure*}

 \begin{figure*}
\begin{center}
\epsfig{file=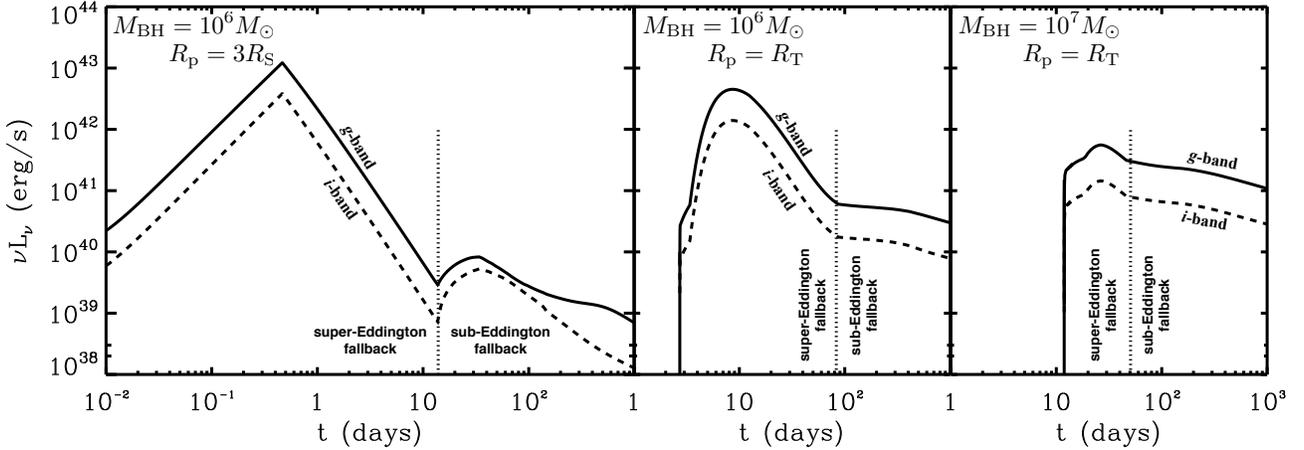, width=18cm}
\vspace{-0.4cm}
\end{center}
\caption{Light curves at $g$- (solid) and $i$-band (dashed) for our
  three fiducial models.  In each panel, times to the left of the
  vertical dotted line ($\sim t_{\rm Edd}$; eq. [\ref{tEdd}]) have
  super-Eddington fallback rates and  an outflow  dominates
  the optical emission; at later times, the fallback rate is
  sub-Eddington and the emission is produced by the accretion disk and
  the photoionized unbound material (see \S\ref{disk+UB emission}).
  In the leftmost panel, the emission rises at early times because the edge of the   outflow  limits the size of the photosphere.  The optical emission then
  declines until the end of the outflow phase as the photosphere recedes and the photospheric temperature rises (Fig. \ref{outflow_spect}).
  \label{outflow_lc}}
\end{figure*}

Particles having the maximum vertical velocity $\Delta \vp$ have
vertical specific angular momentum $j_z = \rp\vp$, and total specific
angular momentum
\begin{equation}
j \sim \rp\vp\left[1+\frac{1}{2}\left(\frac{\Delta \vp}{\vp}\right)^{2}\right] \sim \rp\vp\left[1+\frac{1}{2}\left(\frac{\rstar}{\rp}\right)^{2}\right] 
\end{equation}
to lowest order in $\Delta \vp/\vp$.  The orbital inclination $i$ is
given by $\cos i = j_z/j \sim 1- (1/2)(\rstar/\rp)^2$, so $i \sim
\pm\rstar/\rp$. The resulting inclination dispersion
is $\Delta i \sim 2\rstar/\rp$ (our result is consistent with Evans \&
Kochanek 1989 but we disagree with Khokhlov \& Melia 1996).

The finite inclination dispersion produces a vertical wall of debris
whose inside face scatters, absorbs, and re-radiates a fraction of the disk's emission.  This
face
subtends a solid angle
\begin{eqnarray}\label{dOmega}
\Delta \Omega = \Delta i \Delta \phi & \sim & 48^{1/2}\left(\frac{\rstar}{\rp}\right)^{3/2} \\
& \sim & 0.2 \mbhrat^{-3/2}\rprat^{-3/2}\rstarrat^{3/2} \sr \, . \nonumber
\end{eqnarray}

The number density of particles in the unbound wedge is 
$n \sim (\mstar/2m_p)/(R^2\Delta R\Delta \Omega /3)$,
where $\Delta R$ is the radial dispersion of the material at fixed $\phi$.
This dispersion is due to differences in the particles' radial velocities and azimuthal positions when the star passes through pericenter.  Particles at $\sim R_{\rm max}$ travel on orbits whose 
pericenter is shifted from the star's pericenter by an angle $\sim\pm3(\rstar/\rp)$, which produces a spread in radial position $(\Delta R/R)_{\rm max} \sim (3\rstar/\rp)^{1/2}$.  The number density is then
\begin{equation} \label{density}
n \sim 10^9 \mbhrat^{7/2}\rprat^5\mstarrat\rstarrat^{-7/2}\trat^{-3}\cm^{-3}
\end{equation}
and the radial column density seen by the black hole is
\begin{equation} \label{coldensity}
N \sim 10^{25} \mbhrat^{5/2}\rprat^{7/2}\mstarrat\rstarrat^{-5/2}\trat^{-2} \cm^{-2} \, .
\end{equation}

As the unbound material expands, it cools very quickly; after at most a few weeks, the gas would all be neutral if not for the disk's ionizing radiation.  This radiation ionizes the surface layer of the unbound material.  The ionized gas in turn emits via bremsstrahlung, radiative recombination, and lines.  The physical conditions and processes here are similar to those in the broad line region of an active galactic nucleus (AGN).

The ionized gas can reach photoionization equilibrium provided
conditions change more slowly than the hydrogen recombination rate
$t_{\rm rec}^{-1} \sim n_{\rm e} \alpha_{\rm rec}$.  The recombination
coefficient for hydrogen is $\alpha_{\rm rec} \approx
4\times10^{-13}\cm^3\s^{-1}$, and $n_{\rm e}$ is the electron number
density.  In the ionized region, $n_{\rm e}/n \approx 1$, as we show
below.  The material can remain in equilibrium for at least a few
years, until $t_{\rm rec}/t \gtrsim 1$:
\begin{equation}
\frac{t_{\rm rec}}{t} \sim (n\alpha_{\rm rec} t)^{-1} \sim 10^{-3} \mbhrat^{-7/2}\rprat^{-5}\mstarrat^{-1}\rstarrat^{7/2}\trat^{2} \, .
\end{equation}
The column depth of the ionization front is $N_{\rm ion} \sim 10^{23}U\cm^{-2}$, where
$U  \equiv  L_{\rm disk}/4\pi \rmax^2c\langle h\nu\rangle n$ is the ionization parameter,
\begin{equation}
U\sim 
\end{equation}
\begin{displaymath}
0.3 \left(\frac{L_{\rm disk}}{L_{\rm Edd}}\right)\hnurat^{-1}\mbhrat^{-3/2}\rprat^{-3}\mstarrat^{-1}\rstarrat^{5/2}\trat \, . \nonumber
\end{displaymath}
The electron density in the ionized layer is $n_{\rm e} \approx n(1-10^{-6}U^{-1}) \approx n$ and
the fractional depth of the ionization front is
\begin{displaymath}
\frac{\Delta R_{\rm ion}}{\Delta R}  =  \frac{N_{\rm ion}}{N} \sim \nonumber
\end{displaymath}
\begin{equation}
3\times10^{-3} \left(\frac{L_{\rm disk}}{L_{\rm Edd}}\right)\hnurat^{-1}\mbhrat^{-4}\rprat^{-13/2}\mstarrat^{-2}\rstarrat^{5}\trat^3 \, ,\nonumber
\end{equation}
so the ionized layer is typically thin and highly ionized.

\section{Predicted Emission}
\label{emissionprops}

We use the results of \S\S\ref{bound} and \ref{unbound} to calculate
the emission due to the tidal disruption of a solar-type star as a
function of time and wavelength.  We consider a solar-type star
because the stellar mass-radius relation and typical stellar mass
functions imply that these stars should dominate the event rate.  The
two key parameters we vary are the star's pericenter distance $\rp$
and the BH mass $\mbh$.  We consider the mass range $\mbh \sim
10^5-10^8M_\odot$.

\subsection{Super-Eddington Outflows}
\label{outflow_emission}

Early on, when $\dot{M}_{\rm fallback} \gtrsim \dot{M}_{\rm Edd}$ ($t
< t_{\rm Edd}$), outflowing gas likely dominates the emission.  We
calculate its properties using results from \S\ref{outflows}.  

In Figure \ref{outflow_spect} we plot the spectral energy distribution
at various times during the outflow phase, for three fiducial models:
$\mbh = 10^6 M_\odot$ and $\rp = 3\rs$; $\mbh=10^6 M_\odot$ and
$\rp=\rt$; and $\mbh=10^7 M_\odot$ and $\rp=\rt$.  We take nominal
values of $f_v = 1$ and $f_{\rm out} = 0.1$; we discuss the
uncertainties in these parameters in \S\S\ref{outflow_rates_section}
and \ref{discussion}.  The photosphere lies well inside the edge of
the outflow at all times shown in Figure \ref{outflow_spect}.  The
emission from the outflow has a blackbody spectrum, initially peaking
at optical/UV wavelengths.  As time passes and the density
of the outflow subsides, the photosphere recedes and the emission
becomes hotter but less luminous.

\begin{figure}
\centerline{\epsfig{file=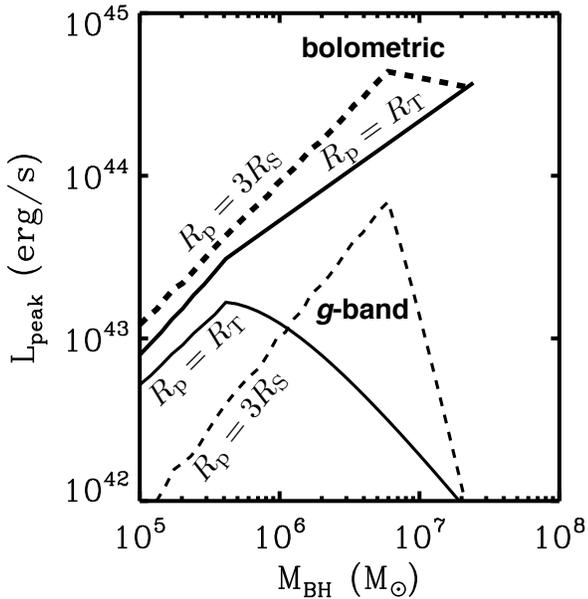, height=9cm}}
\vspace{-0.4cm}
\caption{Peak bolometric (heavy lines) and $g$-band (light lines)
  luminosities of the early-time super-Eddington outflows as functions
  of $\mbh$, for $\rp=3\rs$ (dashed) and $\rp=\rt$ (solid), assuming that
  $f_{\rm out} = 0.1$ and $f_v = 1$.  Figure
  \ref{outflow_tevent} shows the duration of this phase.
  \label{Lpeakfig}}
\end{figure}

In Figure \ref{outflow_lc}, we plot $g$- (4770\AA) and $i$- (7625\AA)
band light curves for the three fiducial models.  In the leftmost
panel at $t \lesssim 1\,{\rm day}$, the edge of the outflow limits the
size of the photosphere, so the photosphere initially expands,
following the edge of the outflow.  After a time $t_{\rm edge}$
(eq. [\ref{tedge}]), however, the photosphere begins to recede inside
the edge of the outflow and the luminosity declines.  In the middle
and rightmost panels, the photosphere lies well inside $R_{\rm edge}$
for virtually the entire outflow phase.  The optical emission
decreases as the photosphere's emitting area decreases and the
temperature rises only slowly.  As Figures \ref{outflow_spect} \&
\ref{outflow_lc} demonstrate, the peak optical luminosity of the
outflow is substantial, $\sim 10^{43} \, {\rm \erg \, s^{-1}} \sim
10^9 L_\odot$, comparable to the optical luminosity of a supernova.
The color of the emission is quite blue ($g-r \approx -0.8$).  To
illustrate how the peak luminosity depends on the parameters of the
disruption event, Figure \ref{Lpeakfig} shows the peak bolometric and $g$-band
luminosities of the outflow as a function of $\mbh$, for
$\rp=3\rs$ and $\rp=\rt$.  For sources at cosmological distances
(which {\it are} detectable; \S \ref{rates}), the negative
k-correction associated with the Rayleigh-Jeans tail implies that the
rest-frame $g$-band luminosity in Figure \ref{Lpeakfig} underestimates
the peak optical luminosity visible at Earth.

\subsection{Disk and Photoionized Unbound Debris}\label{disk+UB emission}

When $\dot{M}_{\rm fallback} \gtrsim \dot{M}_{\rm Edd}$, a fraction of
the falling-back gas is blown away while the remainder likely accretes
via an advective disk (\S\ref{diskmodel}).  As the fallback rate
declines below Eddington, the photons are able to diffuse out of the
region close to the BH and the disk cools efficiently, but also
becomes less luminous.  The vertical dotted line in Figure
\ref{outflow_lc} delineates the super-Eddington fallback (and outflow)
phase from the sub-Eddington fallback phase.  

The accretion disk irradiates the surface of the equatorial unbound
stellar material (\S\ref{unbound}).  In this section we calculate the
combined emission produced by the accretion disk and the irradiated stellar
debris.  In order to isolate the more theoretically secure emission by
the disk and photoionized material, we do {\it not} consider the
emission from super-Eddington outflows in this section.  We show
results for the disk and photoionized material at both $t < t_{\rm
  Edd}$ and $t > t_{\rm Edd}$; depending on the geometry of the
outflow, and the viewing angle of the observer to the source, it is
possible that all three emission components could be visible at early
times.  Because the mass driven away by outflows during the
super-Eddington phase can also be photoionized by the central source
at times $t > t_{\rm Edd}$, our emission line predictions are likely a
lower limit to the total emission line fluxes (\S \ref{discussion}).

\begin{figure*}
\begin{center}
\epsfig{file=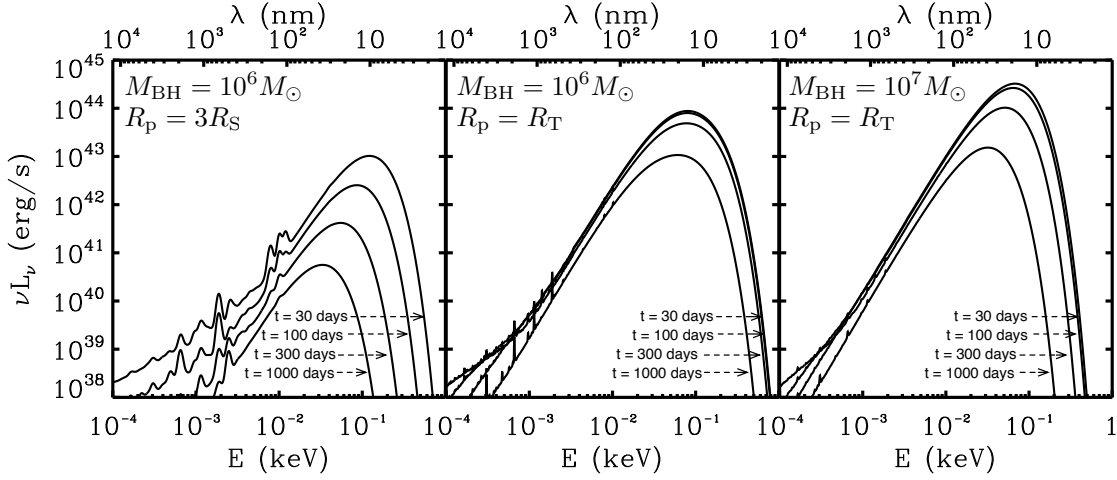, width=16cm}
\vspace{-0.4cm}
\end{center}
\caption{Spectral energy distributions for tidal flares
around a non-rotating BH, 30, 100, 300, and 1000 days after disruption.  Emission from the accretion disk dominates at short wavelengths.  The photoionized unbound stellar debris absorbs and re-radiates some of the disk's emission, producing optical-infrared emission.   These spectra do not include the emission from super-Eddington outflows at early times; see Figs. \ref{outflow_spect} \& \ref{outflow_lc} for this emission.
\label{sed_fiduc}}
\end{figure*}

\begin{figure*}
\begin{center}
\epsfig{file=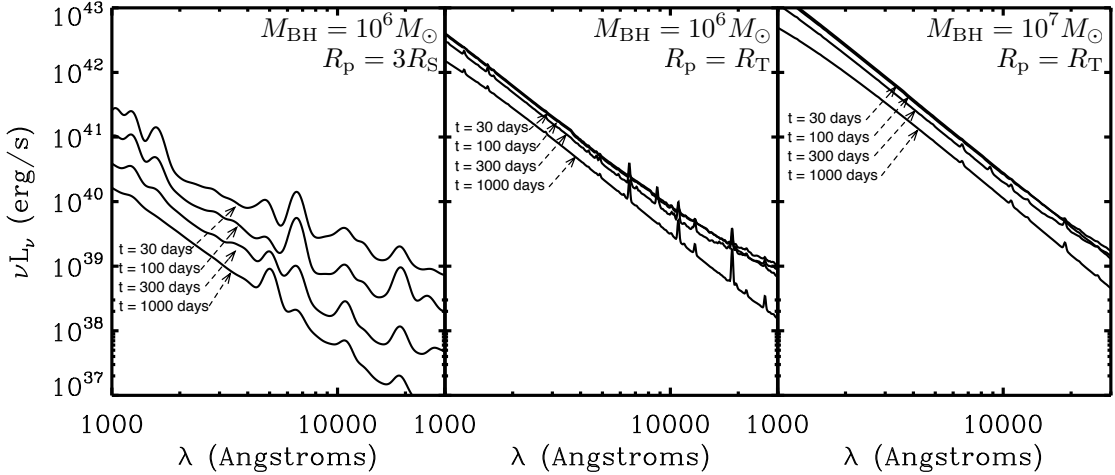, width=16cm}
\vspace{-0.4cm}
\end{center}
\caption{UV to near-infrared spectra for tidal flares around a
  non-rotating BH, 30, 100, 300, and 1000 days after disruption.  The
  spectra are the sum of contributions from the accretion disk and the
  photoionized unbound material, but do not include the emission from
  super-Eddington outflows, which likely dominate at early times
  (Fig. \ref{outflow_lc}).  The linewidths and line strengths are both
  larger for smaller $\rp/\rs$ and smaller $\mbh$
  (eq. [\ref{vmax}]).\label{optIR_fiduc}}
\end{figure*}

We calculate the photoionization properties of the unbound material
using version 07.02.02 of the publicly available code Cloudy, last
described by \citet{ferland}.  We simplify the geometry: the unbound
spray traces out a widening spiral shape with most of the area coming
from close to $\sim \rmax$, so we approximate it as a cloud of area
$\rmax^2\Delta\Omega$ located a distance $\rmax$ from the ionizing
source.  Our model cloud has constant density $n$
(eq. [\ref{density}]), column depth $N$ (eq. [\ref{coldensity}]), and
is irradiated by the accretion disk having the luminosity and spectrum
described in \S\ref{diskmodel}.  The total emission calculated here is
the sum of the emission from this photoionized layer and the emission
from the central accretion disk.  We focus on non-rotating BHs
($R_{\rm LSO} = 3\rs$), although we quote results for rapidly rotating
holes ($R_{\rm LSO}=\rs$) as well.

In Figure \ref{sed_fiduc} we plot the spectral energy distribution 30
days, 100 days, 300 days, and 1000 days after disruption, for our three
fiducial models: $\mbh = 10^6 M_\odot$ and $\rp = 3\rs$; $\mbh=10^6
M_\odot$ and $\rp=\rt$; and $\mbh=10^7 M_\odot$ and $\rp=\rt$.  The
early-time short-wavelength peaks at $\sim 0.1 \, {\rm keV}$ with
luminosity $\sim L_{\rm Edd}$ are emission from the disk.  After a
time $t_{\rm Edd}$, the mass fallback rate declines below the
Eddington rate, and the disk begins to cool and fade.  For $\rp \sim
R_{\rm LSO}$ and $\mbh \sim 10^5 - 10^6M_\odot$, the optical light is
dominated by lines and continuum from the photoionized material.  For
larger $\mbh$ (and larger $\rp/\rs$), the equatorial debris subtends a
smaller solid angle (see eq. [\ref{dOmega}]) and the disk's luminosity
is larger, so the disk dominates the optical emission.

\begin{figure*}
\begin{center}
\epsfig{file=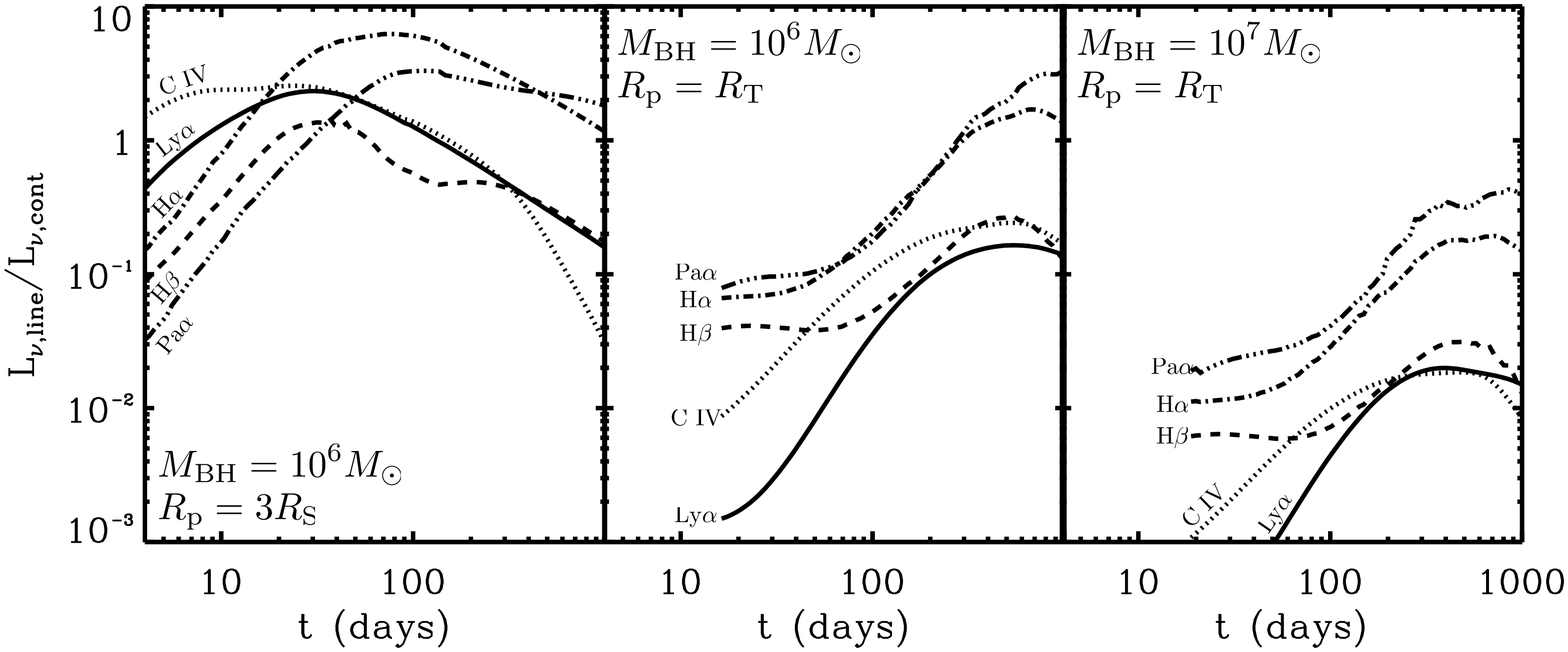, width=16cm}
\vspace{-0.4cm}
\end{center}
\caption{Line strength evolution for the three fiducial models.  Results are shown for Ly$\alpha$ (solid), CIV 1548+1551 (dotted), H$\beta$ (dashed), H$\alpha$ (dot-dashed), and Pa$\alpha$ (triple-dot-dashed).  The quantity $L_{\nu, {\rm line}}/L_{\nu, {\rm cont}}$ is the ratio at line center of the line intensity 
to continuum intensity, taking into account emission from both the disk and the photoionized material, but not the early-time super-Eddington outflows.
\label{lines_fiduc}}
\end{figure*}

\begin{table*}
\caption{Assumed parameters for transient surveys$^{\ref{surveyfoot}}$ and predicted rates.  Our results can be scaled to other survey parameters and model
  assumptions using eq. (\ref{ratescaling}).
  \label{surveytable}}
\begin{tabular}{cccccc}
\hline
\hline
Survey &
$f_{\rm lim}$ &
$f_{\rm sky}$ & 
Cadence &
Rate: D+UB$^{a}$ $(\yr^{-1})$ &
Rate: Outflows$^{b}$ $(\yr^{-1})$ \\
\hline
Pan-STARRS $3\pi$ Survey & 23 AB mag ($g, i$-band) & 0.75 & 6 months & $4-12$ & 200 \\
Pan-STARRS Medium Deep Survey (MDS) & 25 AB mag ($g$-band) & $2\times10^{-3}$ & 4 days & $0.2-1$ & 20 \\
Palomar Transient Factory (PTF) & 21 AB mag ($g$-band) & 0.2 & 5 days & $0.3-0.8$ & 300 \\
Large Synoptic Survey Telescope (LSST) & 24.5 AB mag ($g$-band) & 0.5 & 3 days & $60-250$ & 6000 \\
Synoptic All-Sky Infrared Survey (SASIR)$^c$ & 23.5 AB mag ($Y$-band) & 0.03? & 10 days? & $0.1-0.5$ & 100 \\
ROSAT All-Sky Survey$^d$ & $2\times10^{-12}\erg\s^{-1}\cm^{-2}$ & 1 & $e$ & $2-100$ & N/A$^f$ \\
GALEX Deep Imaging Survey & 25 AB mag ($2316\,{\rm\AA}$) & $7\times10^{-5}$ & $e$ & $0.05-0.2$ & N/A$^f$ \\
\hline
\end{tabular}

$^a$ Rates for the emission from the disk and unbound equatorial debris; the range corresponds to $R_{\rm LSO} = 3 \rs$ (low), $R_{\rm LSO} = 1 \rs$ (high).  This emission is relatively faint in the optical/infrared and may be difficult to detect relative to the host bulge (see \S \ref{discussion}).\\
$^b$ Rates for the emission from super-Eddington outflows, restricted to $z < 1$, for $f_{\rm out} = 0.1$ and $f_v = 1$ (see Fig. \ref{outflowratesfig}). \\
$^c$ The survey strategy for SASIR has not yet been finalized. \\
$^d$ These parameters are for comparison with the all-sky rate calculation by \citet{donley}. \\
$^e$ ROSAT and GALEX do not have regular cadences. \\
$^f$ ROSAT and GALEX have insufficient cadence and/or sky coverage to detect flares from super-Eddington outflows.

\end{table*}

Figure \ref{optIR_fiduc} zooms in on the UV/optical/near-infrared spectra for our three fiducial models.
The emission lines characteristic of the broad line region of an AGN
are typically the strongest lines here as well: e.g., Ly$\alpha$, CIV
1548+1551, H$\beta$, and H$\alpha$.  In most cases, these lines are
optically thick for more than a year.  The lines are extremely broad,
since the marginally bound gas has a speed close to zero while the most
energetic gas leaves the BH at $v_{\rm max} \sim 0.4c$, $0.09c$, and
$0.2c$ for $\mbh =10^6M_\odot$, $\rp=3\rs$; $\mbh = 10^6M_\odot$,
$\rp=\rt$; and $\mbh=10^7M_\odot$, $\rp=\rt$, respectively (see
eq. [\ref{vmax}]).  In addition, the mean velocity along our line of
sight will usually be substantial, so the lines should have a large
redshift or blueshift on top of the galaxy's redshift. For clarity, we
plot the spectra with a mean redshift of zero.

Figure \ref{lines_fiduc} focuses on the evolution of five strong
lines, plotting the ratio $L_{\nu, {\rm line}}/L_{\nu, {\rm cont}}$ at
line center for each.\footnote{By comparing the results of different
  versions of Cloudy, we find that the results for line strengths can
  be uncertain by up to factor few.}  The quantity $L_{\nu, {\rm
    line}}$ is the line intensity at line center accounting for the
significant broadening.  As the surface area of the equatorial wedge
grows in time, line luminosities grow until $\dot M_{\rm fallback}
\lesssim \dot{M}_{\rm edd}$ and irradiation by the disk subsides. The
quantity $L_{\nu, {\rm cont}}$ is the continuum intensity, which
includes the contributions of both the disk (blackbody) and the
photoionized unbound material (bremsstrahlung and radiative
recombination)---again, the emission from the super-Eddington outflows
is not included in $L_{\nu, {\rm cont}}$.  The lines remain prominent
for a few years, and are strongest and broadest for small $\mbh$ and
small $\rp/\rs$.  The UV lines are the strongest lines when the
unbound material dominates the continuum (left panel), while the
near-infrared lines are the strongest when the disk dominates the
continuum (middle and right panels).

We next describe the broadband optical evolution of a tidal disruption
event.  Figure \ref{outflow_lc}, also discussed in \S
\ref{outflow_emission}, plots the optical light curve for each
fiducial model, showing the total emission at both $g$- and
$i$-bands.  For $t \lesssim t_{\rm Edd}$ (left of the dotted
lines), the emission is dominated by the super-Eddington outflows,
while for $t \gtrsim t_{\rm Edd}$ the emission is dominated by the accretion
disk and photoionized equatorial debris.
Once $t > t_{\rm Edd}$, i.e., $\dot M_{\rm fallback} \lesssim \dot
M_{\rm edd}$, the disk's optical luminosity falls off gently,
approximately as $t^{-5/12}$: although the bolometric luminosity is
declining as $t^{-5/3}$, the optical emission lies on the
Rayleigh-Jeans tail.  Increasing $\rp/\rs$ and/or $\mbh$ increases the
disk's luminosity by up to two orders of magnitude because of the
disk's larger emitting area and/or because $L_{\rm Edd}$ rises.
At all times, the disk emission is quite
blue ($g-r \approx -1$).

For large $\mbh$ and/or large $\rp/\rs$ (middle and right panels in
Fig. \ref{outflow_lc}), the disk outshines the photoionized material
at optical wavelengths, and the light curves and color evolution are
determined by the disk emission alone.  By contrast, for
$\mbh=10^6M_\odot$ and $\rp=3\rs$ (left panel), the photoionized
material's optical line emission is initially an order of magnitude
brighter than the disk.  As the illuminating power of the disk
declines but the unbound debris becomes less dense, different lines
wax and wane.  The significant redshift or blueshift of the unbound
material further complicates the photometry by altering which lines
contribute in which wavebands (again, our figures assume a mean
redshift of zero).  These effects can produce a non-monotonic light
curve and a complicated color evolution, depending on the exact
redshift of the source and the velocity of the equatorial debris.

\section{Predicted Rates}\label{rates}
We use our calculated spectra and light curves to predict the number
of tidal disruption events detectable by observational surveys.  We
focus on an (almost) all-sky optical survey like the Pan-STARRS PS1
$3\pi$ survey, but we also predict results for surveys with more rapid
cadence (e.g., PTF and LSST) and discuss the results of our models
compared to ROSAT and GALEX observations.  Our assumed survey
parameters are listed in Table \ref{surveytable}, along with some of
our results.\footnote{These are intended to be illustrative, and may
  not correspond precisely to the true observational survey
  parameters, although we have attempted to be as accurate as
  possible.\label{surveyfoot}}  Our results can readily be scaled to other surveys using
equation (\ref{ratescaling}) discussed below.

To predict rates, we use the redshift-dependent BH mass function given by \citet{hopkins}.
At $z\sim 0$, the BH density is $\simeq 10^{-2}\,{\rm Mpc}^{-3}$ for $\sim10^6M_\odot$ and
gently falling at higher masses; as $z$ rises to 3, the BH number density falls by $\sim 30$.
We assume that the BH number density for
$10^5 - 10^6M_\odot$ is the same as for $\sim10^6M_\odot$,
although it is poorly constrained observationally.  We do not consider tidal disruption events
beyond $z \sim 3$.
(Our results can easily be scaled to other assumed BH mass densities;
see eq. [\ref{ratescaling}]).

The rate of tidal disruptions within a single galaxy is
$\gamma(\mbh)$.  To predict detection rates, we assume that $\gamma$
is independent of BH mass.  We adopt $\gamma = 10^{-5}\yr^{-1}$ as
found by \citet{donley} using the ROSAT All-Sky Survey, which is also
in line with conservative theoretical estimates.
We further assume that this rate is distributed equally among
logarithmic bins of stellar pericenter distance $\rp$, so that
$d\gamma/d\ln\rp = \gamma/\ln(\rt/R_{\rm p,min})$.  In the limit of $z\ll1$, the equation for the predicted rate is
\begin{equation}
\frac{d\Gamma}{d\ln\mbh} = \int_{R_{\rm p,min}}^{\rt} \frac{4\pi}{3}d_{\rm max}^3 f_{\rm sky}
\frac{dn}{d\ln\mbh} \frac{d\gamma}{d\ln\rp} \, \,  d\ln\rp \, 
\label{rateeqn}
\end{equation}
where $f_{\rm sky}$ is the fraction of sky surveyed; when necessary we
use the generalization of equation (\ref{rateeqn}) that includes
cosmological effects.
When the duration of a flare $t_{\rm flare}$ is shorter
than the cadence of the survey $t_{\rm cad}$, we approximate the
probability of detection as $t_{\rm flare}/t_{\rm cad}$.

We start by considering emission from only the accretion disk and
photoionized equatorial debris.  Then in
\S\ref{outflow_rates_section}, we include the emission from
super-Eddington outflows, where the physics is somewhat less certain,
but the observational prospects are particularly promising.

\subsection{Disk and Photoionized Material}\label{diskub_rates_section}
\begin{figure}
\centerline{\epsfig{file=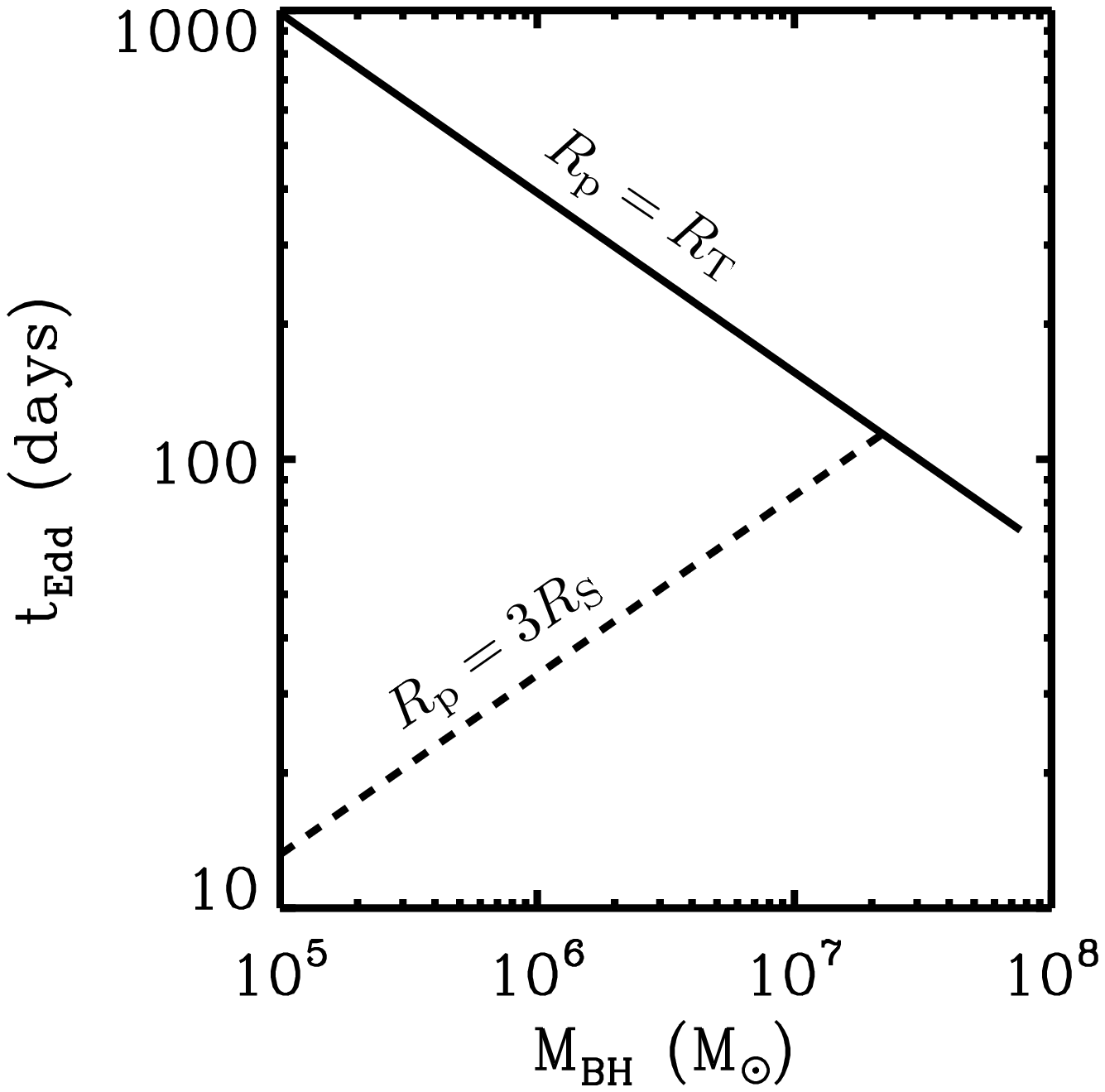, height=9cm}}
\vspace{-0.4cm}
\caption{Duration of maximum luminosity during the late-time accretion
  disk phase,
  plotted as a function of $\mbh$ for $\rp =3\rs$ and $\rp = \rt$.
  Most flares last longer than $\sim$ a month.
  The super-Eddington
  outflow at early times produces a shorter flare that should precede
  this emission (Figs. \ref{outflow_lc} \&
  \ref{outflow_tevent}--\ref{outflowratesfig}).
  (The fallback rates for $\mbh \gtrsim 8\times 10^7\msun$ are never super-Eddington,
  so the flare duration in these systems is $\sim\tfall$.)
  \label{disk_tevent}}
\end{figure}

For all but the largest $\mbh$, the duration of peak optical emission for the accretion disk and
photoionized material is $t_{\rm flare} \sim t_{\rm Edd}$
(eq. [\ref{tEdd}]).  This timescale depends on the BH's mass and the star's
pericenter distance, as shown in Figure \ref{disk_tevent}.  For
$\rp\sim\rt$, the flare lasts for $t_{\rm Edd} \sim 0.3 - 1 \yr$ and
then decays only gently since the disk dominates the optical emission.
However, for $\mbh \sim 10^5 - 10^6 M_\odot$ and $\rp \sim \rlso$, the
optical flare is shorter, $t_{\rm Edd} \sim 0.03 - 0.1 \yr$, and then
the emission decays more quickly since irradiation of the unbound
material---the dominant source of optical emission---subsides.
For $\mbh \gtrsim 8\times 10^7\msun$, $t_{\rm flare} \sim \tfall$ (eq. [\ref{tfallback}])
 because the fallback rate in these systems is never super-Eddington.

\begin{figure*}
\begin{center}
\epsfig{file=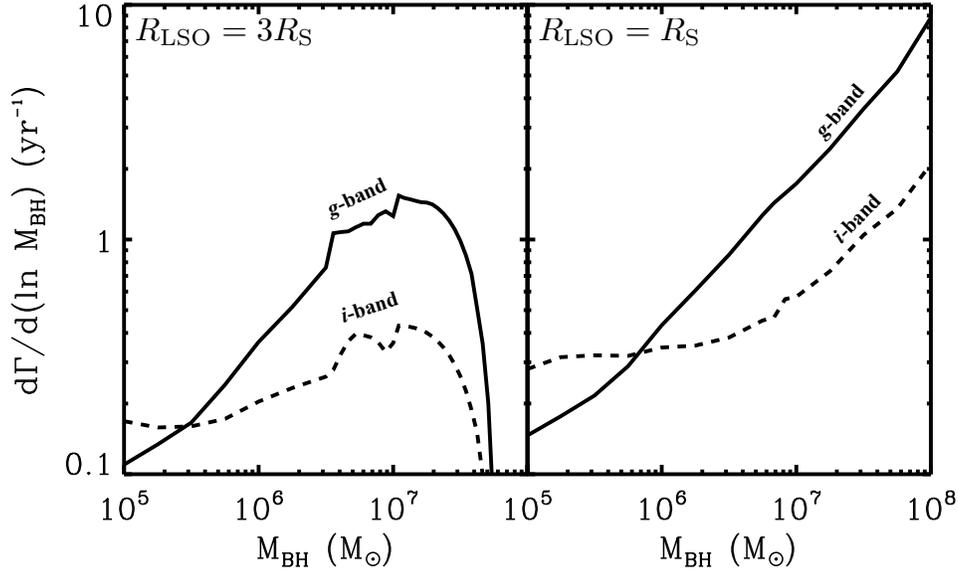, width=14cm}
\vspace{-0.4cm}
\end{center}
\caption{Predicted detection rates as a function of BH mass for the
  Pan-STARRS $3 \pi$ survey (see Table \ref{surveytable}).  Results
  are shown for non-rotating BHs (left panel) and rapidly rotating BHs
  (right panel).  The
rates shown here can be scaled to other surveys and other model
parameters using eq. (\ref{ratescaling}).   These  rates do not include the emission from super-Eddington outflows at early times; see Figs. \ref{outflowratesrpfig} \& \ref{outflowratesfig} for these results.
\label{ratesfig}}
\end{figure*}

Figure \ref{ratesfig} shows our calculated rates for
optically-detected tidal flares (for a survey like the Pan-STARRS
3$\pi$ survey), for both non-rotating and rapidly rotating BHs.  For
$\mbh \gtrsim 10^6 \, M_\odot$, the disk contributes most of the
emitted power, so the rates increase with $\mbh$ as $L_{\rm disk} \sim
L_{\rm Edd}$ increases.  The rates are dominated by $\rp \sim \rt$ and
$\mbh \sim 2\times 10^7M_\odot$ (non-rotating BHs) and $\mbh \sim
10^8M_\odot$ (rapidly rotating BHs).  Since most of the flares that
dominate the rates have relatively long durations
(Fig. \ref{disk_tevent}), imperfect survey cadence only modifies the
detection rates by $\sim 50\%$.  At $10^5-10^6M_\odot$ and small
$\rp$, the photoionized material re-emits a relatively large fraction
of the disk's power in the optical and boosts the detection rates
significantly.

Integrated over $\mbh = 10^6 - 10^8M_\odot$, our estimated rates for
the Pan-STARRS 3$\pi$ survey, assuming non-rotating BHs, are
$4\yr^{-1}$ and $2\yr^{-1}$ in $g$-band and $i$-band, respectively.
The mass range $10^5-10^6M_\odot$ contributes another $0.4\yr^{-1}$
(both $g$-band and $i$-band), assuming
(probably optimistically) that $dn/d\ln\mbh$ and $\gamma$ are the same
at $10^5M_\odot$ as at $10^6M_\odot$.  If the BH is rotating faster,
$\rp/\rs$ can be smaller.  This allows an accretion disk to form for
even $\mbh \sim {\rm few} \times 10^7 - 10^8M_\odot$, widens the disk
for all $\mbh$, and increases the solid angle of the unbound material.
Indeed for $\mbh\sim 10^5M_\odot$ and $\rp \sim \rs$, the unbound
material covers a quarter of the sky!  (At this point $\rs \sim
R_\odot$ so our approximations begin to break down.)  These effects
raise the total predicted rates for rapidly rotating BHs to $\sim 12 \yr^{-1}$.    

\begin{figure}
\centerline{\epsfig{file=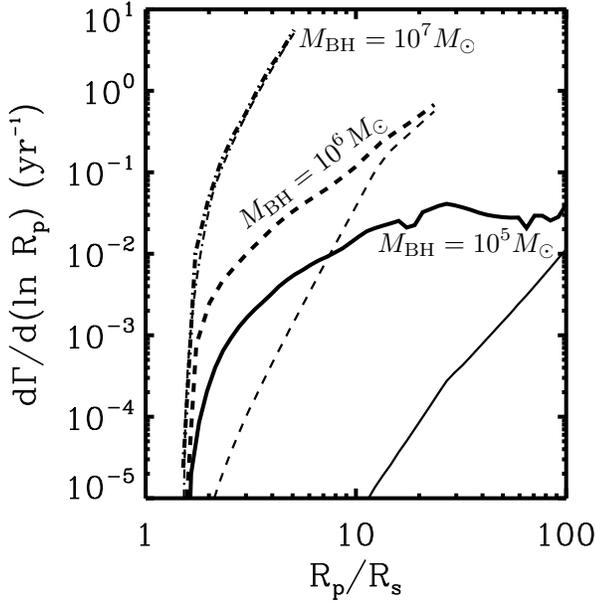, height=9cm}}
\vspace{-0.4cm}
\caption{Predicted detection rates as a function of pericenter distance $\rp/\rs$ for the Pan-STARRS $3 \pi$ survey at $g$-band (see Table \ref{surveytable}). Results are shown for $\mbh=10^5\msun$ (solid), $10^6\msun$ (dashed), and $10^7\msun$ (dot-dashed), assuming a non-rotating BH.  The thin/light lines are for the disk emission alone, while the thick/heavy lines include the emission from both the disk and the photoionized unbound material.  The photoionized material  significantly increases the rates for low $\mbh$ and small $\rp/\rs$.  These rates do not include emission from super-Eddington outflows at early times; those results are shown in Figs. \ref{outflowratesrpfig} \& \ref{outflowratesfig}.  The small fluctuations in the results for $\mbh = 10^5\msun$ are due to difficulties in performing Cloudy
  calculations at very high densities.
  \label{rpbbrates}}
\end{figure}

Figure \ref{rpbbrates} plots the detection rates as a function of
$\rp/\rs$ for $\mbh = 10^5\msun$, $10^6\msun$, and $10^7\msun$, for
the disk alone (light lines) and for the disk plus photoionized material
(heavy lines).  The photoionized material enhances detection rates
significantly for most $\rp/\rs$ at $10^5\msun$ and $10^6\msun$, but
has little effect for $10^7\msun$.  The rates decrease substantially
for $\rp/\rs \rightarrow 1.5$ because then the outer radius of the
disk $\simeq 2 \rp = 3 \rs$ is at the last stable orbit; our disk
model assumes a no-torque boundary condition at $R_{\rm LSO}$,
implying that there is essentially no emission from the disk when $\rp
\sim R_{\rm LSO}$.

Although the rates quoted above and in Figures \ref{ratesfig} and
\ref{rpbbrates} are for a survey covering 3/4 of the sky, assuming
constant $\gamma(M_{\rm BH})$ and constant BH mass density below
$\simeq 10^6 M_\odot$, our predicted rates can be scaled to
other assumed parameters:
\begin{equation}\label{ratescaling}
\frac{d\Gamma}{d\ln\mbh} \propto f_{\rm lim}^{-3/2}f_{\rm sky}f_{\rm cad}\frac{dn}{d\ln\mbh} \gamma(M_{\rm BH})
\end{equation}
assuming that the sources are at $z \ll 1$, where $f_{\rm cad} \equiv \min(t_{\rm flare}/t_{\rm cad}, 1)$.

The Pan-STARRS MDS, PTF, and LSST will have cadences of only a few
days, giving them sensitivity to all events within their survey
volumes.  However, for detecting this relatively long-lived emission,
the advantage of fast cadence is only minor; more significant is the
spatial volume probed by the survey (see Table \ref{surveytable}).
The MDS will deeply ($m_{\rm AB} \sim 25$) image a relatively small
(84-deg$^2$) region, and so will detect $\sim5-10\%$ as many events as
the $3\pi$ Survey, $\sim 0.2-1$ per year.
PTF will image a larger region ($8000 \deg^2$) but less deeply
($m_{\rm AB} \sim 21$), and so will have detection rates similar to
the MDS.
The emission from early-time super-Eddington outflows may
significantly increase these rates, as we discuss in \S
\ref{outflow_rates_section}.  In the next decade, LSST will image a
large region (20,000 ${\rm deg}^2$) deeply ($m_{\rm AB} \sim 24.5$),
and so will discover hundreds of tidal flares.  MDS and LSST will
detect events at cosmological distances ($z_{\rm max} \sim 0.3-0.6$),
where the negative k-correction of the disk's blackbody peak enhances
rates by a factor of a few.  By co-adding images up to $\sim 1 {\, \rm
  month}$, these short-cadence surveys will also be able to raise
their detection rates of events having $\mbh \sim 10^6 - 10^8 M_\odot$
by a factor of a few.  Also in the next decade, SASIR plans to deeply
image 140 ${\rm deg}^2$ each night in the near infrared (to $m_{\rm
  AB} \sim 23.5$ at $Y$-band; \citealt{bloom_sasir}).  Their observing
strategy is not yet finalized; assuming a 10-day cadence covering 1400
deg$^2$ (a good strategy for detecting flares during the
super-Eddington outflow phase; see \S\ref{outflow_rates_section}),
this survey should detect a flare every few years from the accretion
disk plus photoionized debris.

\subsection{Current Observational Constraints}

\label{data}

\begin{figure}
\centerline{\epsfig{file=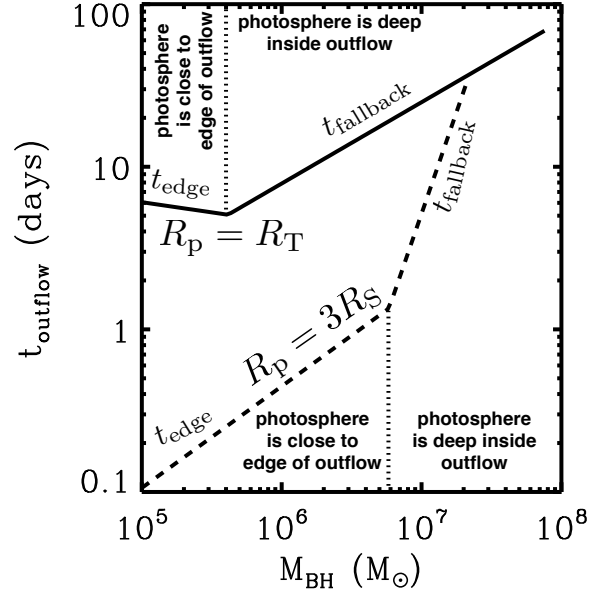, height=9cm}}
\vspace{-0.4cm}
\caption{Duration of peak luminosity during the early super-Eddington
  outflow phase, as a function of $\mbh$ for $\rp =3\rs$ and $\rp=\rt$
  (for $f_{\rm out} = 0.1$ and $f_v = 1$; see eqs. [\ref{mdotout}] \&
  [\ref{vout}]).  The vertical dotted lines mark the boundary between
  events where the edge of the outflow limits the size of the
  photosphere (lower BH masses) and where it does not (higher BH
  masses).  The flares from super-Eddington outflows typically last longer than the few-day   cadence of surveys like PTF, Pan-STARRS MDS, and LSST, but they are often short enough that they would not be detected in surveys
  optimized solely for supernovae.
  \label{outflow_tevent}}
\end{figure}

\begin{figure}
\centerline{\epsfig{file=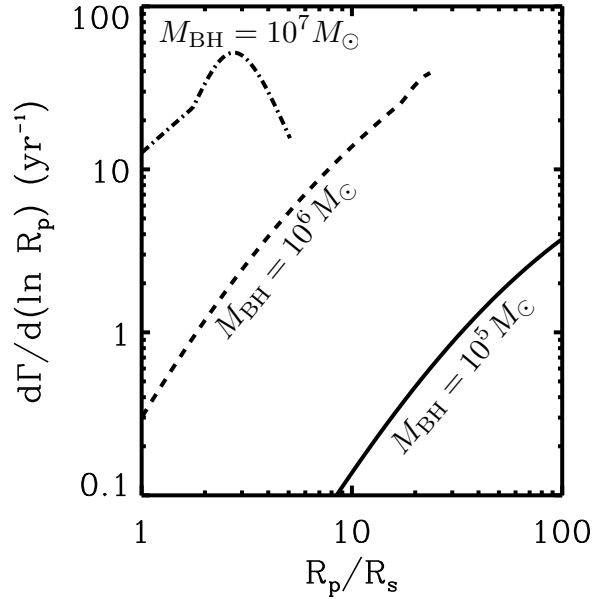, height=9cm}}
\vspace{-0.4cm}
\caption{Predicted detection rates for emission from super-Eddington
  outflows at early times (for $f_{\rm out} = 0.1$ and $f_v = 1$; see
  eqs. [\ref{mdotout}] \& [\ref{vout}]).  Results are shown as a
  function of $\rp/\rs$ for $\mbh=10^5\msun$, $10^6\msun$, and
  $10^7\msun$, all for a Pan-STARRS $3\pi$-like survey at $g$-band
  (see Table \ref{surveytable}).
  \label{outflowratesrpfig}}
\end{figure}

As a check on our model, we have calculated rates for the ROSAT
All-Sky Survey for comparison with \citet{donley}'s result of 42
events per year over the whole sky for $f_{\rm lim} = 2\times 10^{-12}
\erg\s^{-1}\cm^{-2}$ ($0.2 - 2.4$ keV).\footnote{The cadence is $\sim
  1$ year but irregular; we assume it is perfect for simplicity.  The
  rates we predict thus may be slightly high.}  Considering blackbody
emission alone, we predict $2\yr^{-1}$ ($\rin=3\rs$) or $40\yr^{-1}$
($\rin=\rs$).  This strong sensitivity to $\rin$ arises because the
ROSAT band is on the Wien tail of the disk emission.  If we assume
that 10\% of the emission is in an X-ray power-law tail with a photon
index $\Gamma=3$ \citep[not unreasonable assumptions for X-ray
emission from accreting BHs; e.g.,][]{kblaes}, our predicted rates are
$10\yr^{-1}$ ($\rin = 3\rs$) and $100\yr^{-1}$ ($\rin=\rs$).  Given the
large uncertainties in the X-ray emission, our predictions are
consistent with the observational results. We also compare with
detection rates in the GALEX Deep Imaging Survey \citep{gezari08}.
They search an area of 2.882 square degrees, observed at FUV (1539
\AA) and NUV (2316 \AA) down to $f_{\rm lim} \sim 25$ AB magnitudes.
Gezari et al. detect 2 events\footnote{The candidate of
  \citet{gezari09} is in a different field.} in this region over 3
years.  This is somewhat higher than our predicted rates in the NUV,
$0.05\yr^{-1}$ if $\rin=3\rs$ and $0.2\yr^{-1}$ if $\rin=\rs$ (assuming perfect
cadence for simplicity),
and may suggest that the disruption rate per galaxy $\gamma$ is a factor
of a few higher than we have assumed here.

As the above estimates demonstrate, consistency with GALEX and ROSAT
constraints prefers a rate per galaxy of $\gamma \sim (1-3)\times
10^{-5}\yr^{-1}$.  Significantly larger disruption rates, as some
calculations predict (e.g., \citealt{mp}), are inconsistent with
current observational limits unless dust obscuration has significantly
biased the ROSAT and GALEX results or the large disruption rates are
confined to brief epochs in a galaxy's life (e.g., a merger).  Note
also that this constraint only applies to massive BHs with $M_{\rm BH}
\sim 10^{7} M_\odot-10^8M_\odot$, because UV and X-ray surveys select for these
systems (\S\ref{discussion}).

\subsection{Super-Eddington Outflows}\label{outflow_rates_section}

We now calculate optical detection rates accounting for emission from
the luminous but shorter-lived super-Eddington outflows.  We begin by
plotting the duration of peak emission for the outflow phase,
$t_{\rm flare} \sim t_{\rm  outflow}$, in Figure \ref{outflow_tevent} as a function of $\mbh$
for $\rp = 3\rs$ and $\rp=\rt$.  For $\rp=3\rs$ and $\mbh \lesssim
6\times10^6\msun$, the duration of the outflow is set by the time at
which the photosphere recedes inside the edge of the outflow ($t_{\rm
  outflow} \sim t_{\rm edge}$; eq. [\ref {tedge}]) while for larger
$\mbh$, the duration is set by the timescale for the most bound
material to return to pericenter ($t_{\rm outflow} \sim \tfall$;
eq. [\ref{tfallback}]); for $\rp=\rt$, the transition from $t_{\rm
  edge}$ to $\tfall$ occurs at a somewhat lower BH mass of $\mbh
\lesssim 4\times10^5\msun$. Figure \ref{outflow_tevent} shows that
most flares last longer than the few-day cadences of surveys like PTF,
Pan-STARRS MDS, and LSST, but are much shorter than the months-long
cadence of the Pan-STARRS $3\pi$ survey.  ROSAT and GALEX are unlikely
to have detected events during this phase due to insufficient cadence
and sky coverage.

In Figure \ref{outflowratesrpfig}, we plot detection rates as a
function of $\rp/\rs$ for $\mbh=10^5\msun$, $10^6\msun$, and
$10^7\msun$, all for the Pan-STARRS $3\pi$ survey at $g$-band.  At
small $\rp/\rs$, the edge of the outflow limits the size of the
photosphere.  As $\rp/\rs$ increases, the radius of the photosphere at
$t_{\rm edge}$ increases and so the rates increase.  For $\mbh =
10^6\msun$ and $10^7\msun$ at the largest $\rp/\rs$, the photosphere
is no longer limited by the edge of the outflow, and the maximum
luminosity occurs at $\tfall$.  For large $\rp/\rs$, the rate declines
as the photosphere recedes inward.

\begin{figure*}
\begin{center}
\epsfig{file=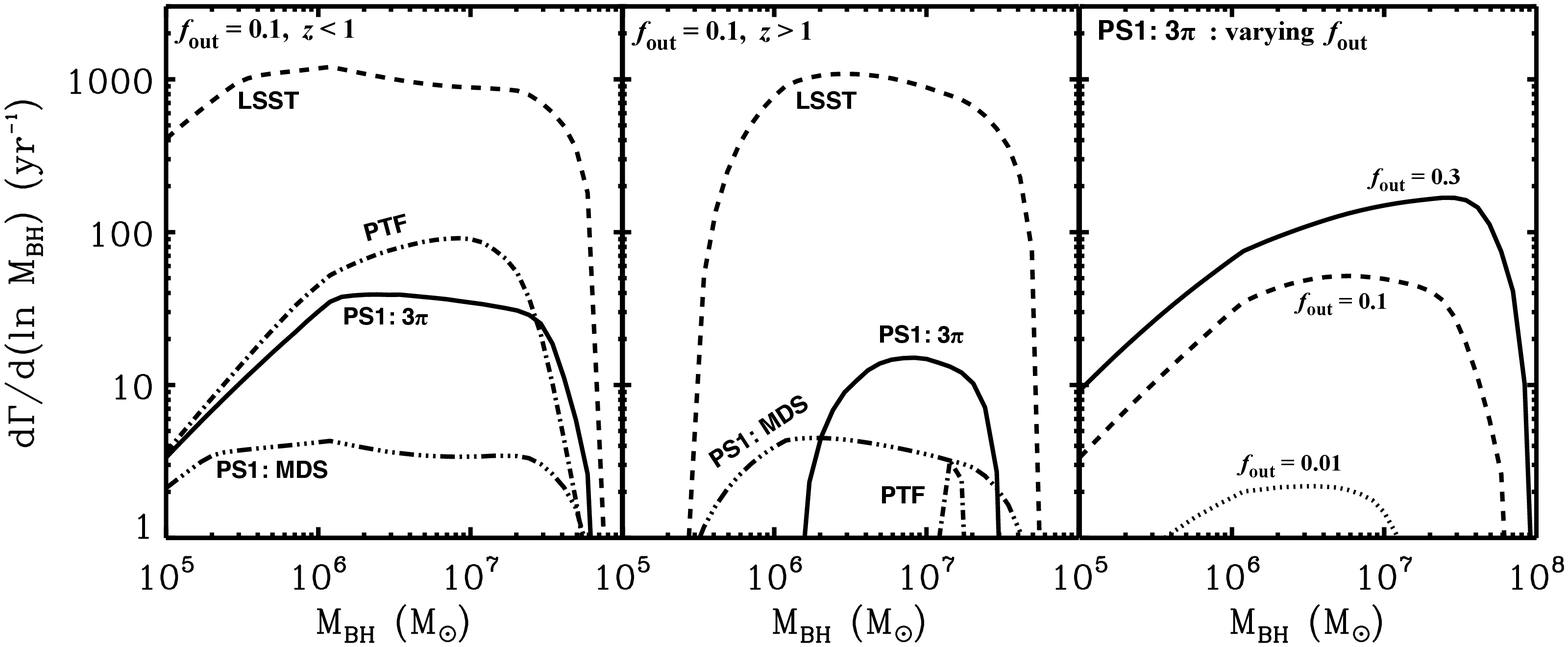, width=16cm}
\vspace{-0.4cm}
\end{center}
\caption{Predicted detection rates for emission at $g$-band from super-Eddington
  outflows for various optical transient surveys.  In the leftmost
  panel, we use our standard outflow model in which $f_{\rm out} =
  0.1$ and $f_v = 1$ (see eqns [\ref{mdotout}] \& [\ref{vout}]), and restrict detections to $z < 1$.
  In the central panel, we plot detection rates for $z > 1$ to illustrate
  the potential for surveys to be sensitive to the cosmological
  evolution of BHs; note that PTF is not deep enough to detect flares beyond $z \sim 1$.
  In the rightmost panel, we vary $f_{\rm out}$ from
  0.3 to 0.01 to illustrate the sensitivity of our results to
  uncertainties in the outflow physics.
  \label{outflowratesfig}}
\end{figure*}

In Figure \ref{outflowratesfig}, we plot overall detection rates at
$g$-band for Pan-STARRS $3\pi$ and MDS, PTF, and LSST.  In the
leftmost panel, we assume that $10\%$ of the falling back material
flows out in the wind ($f_{\rm out}=0.1$), as we have previously.
Here we restrict detections to redshifts $z < 1$, where our assumed
tidal disruption rate per galaxy is appropriate.
The Pan-STARRS $3\pi$ survey should detect $200\yr^{-1}$, while the
MDS should detect $20\yr^{-1}$ because of its smaller spatial volume.
PTF should detect $300\yr^{-1}$ as well, since its fast cadence makes
up for its smaller spatial volume relative to the Pan-STARRS $3\pi$
survey.  LSST's large spatial volume and rapid cadence should allow it
to detect $6000\yr^{-1}$!  Assuming the survey parameters and
strategy described in \S\ref{diskub_rates_section} and Table
\ref{surveytable}, SASIR should detect $\sim 100\yr^{-1}$ as well.

In the central panel of Figure \ref{outflowratesfig},
we plot detection rates for redshifts $z > 1$ to
highlight the possibility of studying tidal disruption events at
cosmological distances.  Deep surveys like MDS will be sensitive to
events far beyond $z \sim 1$.  Even the shallower $3\pi$ survey could
detect $\sim 30$ events per year at $z > 1$.  Our predicted rates at
$z > 1$ are particularly uncertain: higher redshift sources are more
likely to be obscured,
and the lower mass BHs of interest for tidal
disruption are still growing significantly in mass at these redshifts
\citep{heckman} so the tidal disruption rate and mass function become less certain.
Nonetheless, the fact that tidal disruption flares will be detectable
at $z > 1$ with forthcoming surveys highlights that these sources may
become a strong probe of the evolution of $\sim 10^6-10^8 M_\odot$
BHs.  

In the rightmost panel of Figure \ref{outflowratesfig}, we show the
sensitivity of our predictions to uncertainty in the outflow model by
plotting the rates for the Pan-STARRS $3\pi$ survey for different
assumptions about the fraction of the material that is blown away in
the outflow, $f_{\rm out}$.  For $f_{\rm out} \sim 0.3$, the detection
rate is $600\yr^{-1}$, while for $f_{\rm out} \sim 0.01$ it falls to
$8\yr^{-1}$.  The rate falls rapidly at the highest $\mbh$ because
the outflow is optically thin.  Figure \ref{outflowratesfig} shows
that even if we very conservatively assume that only $1 \%$ of the
material is blown away when the fallback rate is super-Eddington,
upcoming optical surveys like Pan-STARRS should still be able to
detect a significant number of tidal disruptions during the
super-Eddington phase.

\section{Discussion}\label{discussion}

We have calculated the spectra and light curves produced by the tidal
disruption of a solar-type star by a massive black hole.  Upcoming
optical transient surveys should detect many such events (\S
\ref{rates}).  Our results demonstrate that there are at least three
different emission components that are important during tidal flares:
(1) outflows at early times when the fallback rate is super-Eddington,
(2) a compact ($\lesssim 10-100 \rs$) accretion disk around the BH,
and (3) stellar debris that is unbound during the disruption and forms
an outflowing ``wedge'' in the equatorial plane (see
Fig. \ref{UBdiagram}).  It is also possible that the super-Eddington
fallback powers a lower-density magnetically-dominated jet, but the
properties of such a jet are difficult to predict so we do not
consider this potential source of emission.

Each of these three components contributes to the total emission from
tidal disruption flares.  At early times, the super-Eddington outflows
likely dominate, producing a few-- to 10--day optical-infrared flare
with a luminosity comparable to that of a supernova
(Figs. \ref{outflow_spect}-\ref{Lpeakfig} \& \ref{outflow_tevent}).
As the fallback rate decreases below the Eddington rate, these
outflows will diminish, revealing the underlying accretion disk that
emits primarily in the UV to soft X-rays (Fig. \ref{sed_fiduc}); at
this time, the optical emission is likely to be much less than that of
a typical AGN---and well below that of the super-Eddington phase
(Fig. \ref{outflow_lc})---because the accretion disk is not very
spatially extended.  The central UV and soft X-ray source photoionizes
the inner edge of equatorial stellar debris, producing a spectum of
broad emission lines (Figs. \ref{sed_fiduc}, \ref{optIR_fiduc}, \&
\ref{lines_fiduc}) whose ``rest'' wavelength should be either
blueshifted or redshifted with respect to the host galaxy depending on
the line of sight of the observer relative to the escaping
material. We find that this spectroscopic signature of tidal flares is
the strongest for low mass BHs because the equatorial stellar debris
occupies the largest solid angle in these systems
(eq. [\ref{dOmega}]).

Although the above stages are the focus of this paper, for
completeness we briefly discuss the rest of the evolution of a tidal
disruption event.  As the fallback rate continues to decrease below
the Eddington rate, the viscous time in the thin disk (eq. [\ref{tvisc}]) increases and
becomes comparable to the time $t$ since disruption;
at this point,
matter begins to build up rather than rapidly accreting onto the BH.
For disruption at $\rp \sim \rt$, which is likely to produce a
significant fraction of the events (Figs. \ref{rpbbrates} \&
\ref{outflowratesrpfig}), we estimate that this occurs $\sim 3 \,
\alpha_{0.1}^{3/7}$ years 
after disruption, when the fallback rate has decreased to
$\dot{M}_{\rm fallback} \sim 0.2 \, \alpha_{0.1}^{-5/7} \,
\mbhrat^{-2/3} \, \dot{M}_{\rm Edd}$ 
(where $\alpha \equiv 0.1 \alpha_{0.1}$ is the dimensionless
viscosity).  Separately, we expect a significant change in the
thermodynamics of the disk when $\dot{M}_{\rm fallback} \simeq
\alpha^2 \dot M_{\rm Edd} \sim 10^{-2} \dot{M}_{\rm Edd}$.  Below this
accretion rate, the material will no longer cool efficiently when it
circularizes and shocks upon returning to pericenter.  Instead of
cooling to form a thin disk, the material will be shock heated to form
a geometrically thick, radiatively inefficient accretion flow.  In
general, both geometrically thin/optically thick and geometrically
thick/radiatively inefficient disks appear to be stable accretion
solutions for a given $\dot M \lesssim 10^{-2} \dot M_{\rm
  Edd}$. However, in this case, the boundary condition that matter
shocks up to the virial temperature upon returning to pericenter picks
out the radiatively inefficient solution once $\dot{M}_{\rm fallback}
\lesssim 10^{-2} \dot M_{\rm Edd}$.  Because the viscous time in a
thick disk is $\sim \alpha^{-1}$ times the dynamical time, once the
accretion becomes radiatively inefficient, {the viscous time is always
  much shorter than the orbital period of matter returning to
  pericenter}.  Moreover, the transition to radiatively inefficient
accretion happens at a time comparable to when matter would otherwise
begin accumulating in a thin disk, particularly for more massive
BHs. This suggests that there is typically only a limited
range of accretion rates (and time) during which the ``spreading
disk'' solution of \citet{cannizzo} applies.  Instead, at late times
matter will rapidly accrete via a thick disk and the accretion rate
will decay as $\sim \dot M_{\rm fallback} \propto t^{-5/3}$.  As in
X-ray binaries \citep{remillard}, we expect that the thermodynamic
transition at $\sim 10^{-2} \dot M_{\rm Edd}$ will be accompanied by a
significant change in the luminosity and spectrum of the disk, and
perhaps also by the production of relativistic jets.  This should be
explored in more detail in future work.

Having summarized our key results and the timeline of a tidal
disruption event, we now discuss some uncertainties in our models,
observational challenges to detecting tidal flares, and the
astrophysical applications of studies of tidal disruption events.

\subsection{Super-Eddington Outflows}
In \S\ref{outflows}, we described our simple model for outflows driven
when the fallback rate is super-Eddington.  Energy conservation
implies that the falling back material initially remains bound after
returning to pericenter and circularizing, but even small amounts of
accretion can release additional energy and drive a powerful outflow.
There is, however, a significant uncertainty in precisely how much of
the falling back material is blown away, and in the kinematics of the
outflow.  We assume that the gas expands roughly spherically from the
BH, but the flow is probably somewhat collimated along the pole, due
to the original angular momentum of the stellar debris.  Some photons
can then leak out through the sides of the outflow rather than
continuing to drive the expansion; in this case, the overall emission
would be somewhat hotter and fainter, with a dependence on viewing
angle.

We have parameterized the terminal velocity of the outflow using
$v_{\rm wind} = f_v v_{\rm esc}(2\rp)$, and the mass outflow rate
using $\dot{M}_{\rm out} \equiv f_{\rm out}\dot{M}_{\rm fallback}$; in
all of our calculations, we have assumed that $f_{v} \sim 1$.  If
the gas actually expands more slowly ($f_v < 1$), its density will be
larger, so the photosphere will be larger, increasing the optical
fluxes and detection rates.  In the extreme case in which there is no
unbound outflow, but super-Eddington fallback leads to a
radiation-pressure-supported atmosphere around the BH that slowly
expands until the photons can diffuse out, we also expect significant
optical luminosities during the super-Eddington phase (e.g.,
\citealt{loeb}).

It is worth noting that our predictions for the radiation from
super-Eddington outflows are particularly uncertain for low $\mbh$ and
small $\rp/\rs$, when the edge of the outflow limits the radius of the
photosphere and determines both the peak luminosity
(Fig. \ref{outflow_lc}) and duration (Fig. \ref{outflow_tevent}) of
the flare; this does not, however, significantly influence our total
predicted rates (Fig. \ref{outflowratesrpfig}).

Figure \ref{outflowratesfig} shows that even if the outflow rate is
just a few percent of the fallback rate ($f_{\rm out} \sim 0.01-0.1$),
the outflowing gas is sufficiently bright in the optical that
forthcoming surveys should detect a significant number of tidal
disruption flares.  This makes early-time optical flares from the
tidal disruption of stars an extremely promising candidate for current
and future optical/infrared transient surveys.  (We discuss some
practical issues associated with detecting these sources in \S
\ref{disc_obs}.)  In this context, we note that in the radiation
hydrodynamic simulations of accretion at $\dot M = 100 \dot M_{\rm
  Edd}$ carried out by \citet{ohsuga}, $\sim 10\%$ of the gas becomes
unbound, $\sim 10\%$ accretes, and the remaining $\sim 80\%$ is
marginally bound and may (or may not) eventually accrete as well.
These precise values will depend on the pericenter of the star, with
smaller $\rp/\rs$ likely leading to smaller $f_{\rm out}$, i.e., a
smaller fraction of the gas being blown away.  Future observational
constraints on the luminosity, spectrum, and timescale of the
super-Eddington outflow phase should be able to strongly constrain the
value of $f_{\rm out}$ in individual events.  These results will have
important implications for how massive BHs grow.  In particular, if
$f_{\rm out}$ is typically modest, this would imply that black holes
can accrete at rates far above the Eddington rate, perhaps helping to
explain how supermassive BHs ($\mbh \gtrsim10^8\msun$) can be
observable as luminous quasars as early as $z \sim 6$.

Figure \ref{outflowratesfig} also demonstrates that deep optical
surveys such as the Pan-STARRS MDS and LSST will be sensitive to tidal
flares at high redshift.  These surveys may thus provide a powerful
probe of the BH mass function and stellar dynamics in galactic nuclei
as a function of redshift.
For example, at $z \sim 0.1$, BHs having $M_{\rm BH} \lesssim 10^7
\msun$ are still growing significantly in mass \citep{heckman} and
thus their disruption rates may evolve significantly with redshift.
In addition, galaxy mergers, which are more common at $z \sim 1-2$,
could substantially increase the tidal disruption rate: \citet{chen}
find rates of up to $\sim 1\yr^{-1}$ for $\sim 10^{5}\,{\rm years}$
after the merger due to three-body interactions between stars and a
binary BH.

\subsection{The Accretion Disk, Photoionized Gas, \& Broad Emission Lines}

We now consider several aspects of our model for the accretion disk,
broad emission lines, and the material unbound during the disruption
(\S\S\ref{diskmodel} and \ref{unbound}).  Our accretion disk model is
designed to describe the emission from the time when the disk first
forms through the following few years.  During much of this period,
the fallback rate is super-Eddington, and we expect the disk to be
optically and geometrically thick, with radiation pressure dominating
gas pressure.  Once the fallback rate becomes sub-Eddington, the disk
becomes geometrically thin and may be subject to viscous instabilities
(although it is thermally stable; \citealt{hirose}).  These
instabilities may lead to additional time dependence not captured in
our models, particularly at late times when the viscous time in the
thin disk becomes comparable to the orbital period of the material
falling back to pericenter.  As described above, once $\dot{M}_{\rm
  fallback} \lesssim \alpha^2\dot{M}_{\rm Edd}$, the density is
sufficiently low that the flow becomes radiatively inefficient and our
model is no longer appropriate.  The disk will then heat up and its
spectrum will become significantly harder.  This phase may be
detectable by hard X-ray transient surveys like EXIST.

It is also unclear how much tidal forces will spin up the rotation of
the star as it approaches pericenter.  We have assumed that the star
is maximally spun up, so that stellar debris is accelerated to
relative velocities $\Delta v \sim \vp(\rstar/\rp)^{1/2}$ in the
azimuthal direction.  If in fact the spin-up is less effective, the
onset of the flare, which occurs at the fallback time
(eq. [\ref{tfallback}]), will be later by a factor of few and the
solid angle subtended by the unbound equatorial debris will be
somewhat smaller.  This will not change our qualitative conclusions,
only some of our quantitative results.

It is important to stress that the line emission we predict may well
be an underestimate in all cases: the accretion disk will also
photoionize the {\it back edge} of the material that was blown away
during the super-Eddington phase, which is far from the BH once the
outflows subside. Simple estimates indicate that the density and
velocity of this outflowing gas are similar to that of the gas unbound
at the time of disruption; as a result, irradiation of this gas will
produce additional broad hydrogen lines.  The equivalent width of
these lines depends on the solid angle subtended by the
super-Eddington winds, which, although uncertain, is likely to be
significant.  These lines are unlikely to depend as sensitively on
$\mbh$ and $\rp/\rs$ as the emission lines from the equatorial
debris (see Fig. \ref{optIR_fiduc} for the latter).  As a result,
observations of the line emission will help constrain the geometry of
the outflowing gas created during the tidal disruption event.

In addition to emission lines from the back edge of the outflowing gas
at late times, the photosphere of the super-Eddington outflow may at early times show
strong {\it absorption lines}, particularly in the ultraviolet (much like the
photosphere of a star); these lines would likely be highly
blueshifted relative to the lines of the host galaxy.  Finally, we
reiterate that we expect very little narrow forbidden line emission
from tidal disruption events: the outflowing stellar debris
is too dense to produce forbidden lines, and there is insufficient time to
photoionize ambient lower density gas far from the BH.
It is possible that there is ambient gas in the galactic nucleus
sufficiently close to the BH to produce forbidden lines on a $\lesssim 1\yr$ timescale,
but the prevalence of such gas is currently poorly understood.

\subsection{Observational Considerations}
\label{disc_obs}

Candidate detections of tidal disruption flares have thus far been
selected by their UV and soft X-ray emission (predominantly via GALEX,
ROSAT and XMM-Newton).  The emission at these wavelengths is primarily
produced by the accretion disk, which is brighter for larger $\mbh$.
As a result, these surveys are most sensitive to BHs having $\mbh \sim
10^7-10^8\msun$.  UV and X-ray selected events are likely to be
discovered somewhat after the initial period of super-Eddington
fallback, because the outflow during that phase probably precludes
direct observation of the underlying accretion disk from many viewing
angles.  This outflowing material could also be a significant source
of obscuration even at late times when the fallback rate is
sub-Eddington.

The accretion disk emission we predict for UV-selected events is
broadly consistent with the GALEX candidates: UV luminosities of $\sim
{\rm few}\times10^{43}\erg\s^{-1}$, optical luminosities
of $\sim{\rm few}\times 10^{41}\erg\s^{-1}$, blackbody
temperatures of $T\sim {\rm few}\times 10^4- {\rm few}\times10^5\K$,
and bolometric luminosities of $L_{\rm bol} \sim 10^{45}\erg\s^{-1}$.
The events selected from soft X-rays have less data, but typically
have soft X-ray luminosities of $\sim 10^{43}- 10^{44}\erg\s^{-1}$.  This
emission may be from the accretion disk, at energies just above the
blackbody peak, or may be from an X-ray power-law tail with $1-10\%$
of the bolometric luminosity.  In addition, as discussed in \S
\ref{data}, our model is consistent with ROSAT and GALEX rate
estimates, provided that the tidal disruption rate per galaxy for BHs
with $M_{\rm BH} \sim 10^{7}M_\odot-10^8 M_\odot$ is $\gamma \sim
10^{-5}\yr^{-1}$.  This constraint already suggests that galactic
nuclei in the nearby universe are relatively spherical, rather than
triaxial, because the expected disruption rate is significantly higher
in the latter case \citep{mp}.

As described in \S\ref{emissionprops}, the spectral signature of the
equatorial stellar debris is a transient spectrum of broad emission
lines shifted in wavelength relative to the host galaxy.  We do not
expect forbidden lines (e.g. [NII], [SII], [OI], [OIII]) to be
present, because the density in the unbound material is too
high.\footnote{The $5007 {\,\rm \AA}$ line of [OIII] does appear at
  late times for the lowest $\mbh$ and $\rp$, where the densities fall
  below this ion's critical density after about a year.}
For several reasons, it is not surprising that this spectral signature
has yet to be seen.  First, the tidal flare candidates are likely from
relatively high-mass BHs; in those events, the unbound stellar debris
subtends a small solid angle (eq. [\ref{dOmega}]) and so the emission
lines should be $\lesssim 1\%$ of the bulge luminosity.  Future
optically-selected tidal flares are more likely to show detectable
lines.  It is also important to note that most standard searches for
AGN in optical/infrared surveys use forbidden lines to identify
nuclear activity, and have not specifically looked for faint, broad
lines offset from the host galaxy's lines.  Tidal disruption events
may yet be hiding in archival spectroscopic
data.  

We predict that outflows during the super-Eddington fallback phase
have peak optical luminosities of $\sim 10^{43}-10^{44} \erg\s^{-1}$
and characteristic decay timescales of $\sim 10$ days
(Figs. \ref{Lpeakfig} \& \ref{outflow_tevent}). These events are
sufficiently bright that a natural concern is whether our predictions
can already be ruled out by optical supernova searches such as the
Supernova Legacy Survey and Stripe 82 in the Sloan Digital Sky Survey.
Although a careful search of archival data is clearly warranted, we do
not believe that current observations are necessarily that
constraining, for two reasons.  First, the outflow phase can be
relatively brief and many survey cadences may be insufficient to find
these events.  Most importantly, however, tidal flares could be
readily mistaken for AGN and thus discarded. Indeed, most supernova
searches discard galactic nuclei in order to avoid confusion with AGN
and optimize their probability of detecting supernovae.

Optically detecting a tidal flare may require disentangling the flare
emission from that of the BH's host galaxy.  For example, at a
distance of 300 Mpc, a ground-based optical survey with a resolution
$\sim 1$" should just be able to resolve a kiloparsec-sized bulge.
Bulges are found to be $\sim 700$ times more massive than their
central BHs \citep{haringrix}.  Super-Eddington outflows typically
shine at $10^{43}-10^{44} \erg\s^{-1}$ (see Fig. \ref{Lpeakfig}), so
this phase should be at least as bright as the host galaxy and readily
detectable given sufficient attention to sources in galactic nuclei
and careful screening to rule out an unsteady AGN.  By contrast, the
typical optical luminosity of the accretion disk itself is $10^{40} -
{\rm few} \times 10^{41} \erg\s^{-1}$ (see Fig. \ref{outflow_lc}).
The accretion disk would thus brighten the host bulge by only a few
percent.  Photometric detections of late-time flares will require very
careful bulge subtraction.  As a result, shallow, wide-area surveys
such as PTF are more likely to find the late-time disk emission than
narrow deep surveys such as the Pan-STARRS MDS.  As an additional
complication to finding tidal flares, optical extinction in galactic
nuclei can be significant (although less than at UV or soft X-rays);
as a result, some fraction of optical tidal flares may not be
detectable due to obscuration.  Infrared surveys such as SASIR, which
are also very sensitive to tidal disruption events (Table
\ref{surveytable}), will be particularly immune to the effects of
obscuration.

Type II supernovae in the nuclear regions of galaxies may be confused
with tidal disruption events, as both have quite blue colors. For
sources at $\sim 300$ Mpc, we estimate that such supernovae will occur
within $\sim 1"$ of the galactic nucleus at a rate of $\sim
10^{-4}\yr^{-1}$, perhaps an order of magnitude more often than tidal
disruption events; at higher redshift the contamination from
supernovae will be more significant, but follow-up imaging at high
spatial resolution and/or spectroscopic follow-up should help classify
these events and distinguish tidal flares from nuclear supernovae.

\subsection{Astrophysical Applications}

Theoretical calculations of the tidal disruption rate per galaxy, $\gamma$, vary
substantially and can have complicated dependences on BH mass and
pericenter distance \citep[e.g.,][]{magtrem}.  The rate model we implement is consistent with
theoretical estimates and is sufficiently simple that the reader can
easily scale our results to different model parameters
(eq. [\ref{ratescaling}]).  
We have assumed that the rate at which stars enter the disruption zone
($\rs < \rp < \rt$) is independent of BH mass, and constant with $\ln
\rp$.  A star may venture deep into the disruption zone ($\rp
\sim \rs$) on its last orbit if its change in angular momentum over
one dynamical time is large enough---at least of order the maximum
angular momentum for disruption.  This condition is satisfied in the
full loss cone regime, and marginally satisfied in the outskirts of
the diffusive regime.  For realistic stellar density profiles, the
disruption rate is dominated by
the boundary between these two regimes \citep[e.g.,][]{alex05}, so
many stars probably do take large enough angular momentum steps to
arrive at $\rp \ll \rt$.  In the diffusive regime, the disruption rate
per $\ln \rp$ varies weakly with $\ln \rp$, consistent with our
assumption in \S \ref{rates}.
Given the sensitivity of the optical-infrared emission from tidal flares to
$\rp$, upcoming surveys should significantly improve our knowledge of
the stellar dynamics in galactic nuclei.

Our results demonstrate that optical transient surveys will be quite
sensitive to the lowest mass BHs in galactic nuclei, both because of
the outflows produced when the fallback rate is super-Eddington and
because of the large angle subtended by the equatorial stellar debris.
Such BHs are otherwise difficult to detect because their host galaxies
are faint, it is difficult to resolve their small spheres of
influence, and even when they are active, their Eddington luminosities
are low.  The space density of $10^5-10^6M_\odot$ BHs and the stellar
density profiles in the galaxies they inhabit are only moderately
well-constrained at present \citep[e.g.,][]{greene07}, as is the role
of tidal disruption in growing these BHs. Optical searches for tidal
flares should thus prove to be a powerful probe of low-mass BHs and
their host galaxies.

\section*{Acknowledgments}
We thank Josh Bloom, Phil Chang, Eugene Chiang, Luis Ho, and Enrico
Ramirez-Ruiz for helpful discussions; we also thank Phil Hopkins for providing the black hole mass function
in tabular form; and we thank Gary Ferland, Ryan Porter and Peter van Hoof
at the Cloudy discussion board for rapid and helpful replies to our questions.  LES and EQ are
supported in part by the David \& Lucile Packard Foundation and NASA
Grant NNG06GI68G.  LES dedicates this work to the memory of two
shining stars, Sandra Strubbe and Doug Baker.

\label{lastpage}


\begin{thebibliography}{}
\bibitem[Abramowicz et al.(1988)]{abram} Abramowicz M.A., Czerny B., Lasota J.P., Szuszkiewicz E. 1988, \apj, 332, 646
\bibitem[Alexander(2005)]{alex05} Alexander T. 2005, Physics Reports, 419, 65
\bibitem[Ayal et al.(2000)]{ayal} Ayal S., Livio M., Piran, T. 2000,
  \apj, 545, 772
\bibitem[Blandford \& Begelman(1999)]{blandfordbegelman} Blandford, R.D., Begelman M.C. 1999, \mnras, 303, L1
\bibitem[Bloom et al.(2009)]{bloom_sasir} Bloom J.S., Prochaska J.X., Lee W. et al. 2009, arXiv:0905.1965v1 [astro-ph.IM]
\bibitem[Bogdanovi\'c et al.(2004)]{bogdanovic} Bogdanovi\'c T., Eracleous M., Mahadevan S., Sigurdsson S., Laguna P. 2004, \apj, 610, 707
\bibitem[Brassart \& Luminet(2008)]{brassart} Brassart M., Luminet J.-P. 2008, \aa, 481, 259
\bibitem[Cannizzo et al.(1990)]{cannizzo} Cannizzo J.K., Lee H.M., Goodman J. 1990, \apj, 351, 38 
\bibitem[Chen et al.(2009)]{chen} Chen X., Madau P., Sesana A., Liu, F.K. 2009, arXiv:0904.4481v1 [astro-ph.GA]
\bibitem[Donley et al.(2002)]{donley} Donley J.L., Brandt W.N., Eracleous M., Boller T. 2002, \apj, 124, 1308
\bibitem[Esquej et al.(2007)]{esquej} Esquej P., Saxton R.D., Freyberg M.J., Read A.M., Altieri B., Sanchez-Portal M., Hasinger G. 2007, A\&A, 462, L49
\bibitem[Evans \& Kochanek(1989)]{evkoch} Evans C., Kochanek C. 1989, \apjl, 346, L13
\bibitem[Ferland et al.(1998)]{ferland} Ferland G.J., Korista K.T., Verner D.A., Ferguson J.W., Kingdon J.B., Verner E.M. 1998, \pasp, 110, 761
\bibitem[Gezari et al.(2006)]{gezari06} Gezari S., Martin D.C., Milliard B. et al. 2006, \apj, 653, L25
\bibitem[Gezari et al.(2008)]{gezari08} Gezari S., Basa S., Martin D.C. et al. 2008, \apj, 676, 944
\bibitem[Gezari et al.(2009)]{gezari09} Gezari S., Heckman T., Cenko S.B. et al. 2009, arXiv:0904.1596v1 [astro-ph.CO]
\bibitem[Greene \& Ho(2007)]{greene07} Greene J.E., Ho L.C. 2007, \apj, 670, 92
\bibitem[Grindlay(2004)]{grindlay} Grindlay J.E. 2004, AIP Conf. Proc., 714, 413
\bibitem[Guillochon et al.(2008)]{guillochon} Guillochon J., Ramirez-Ruiz E., Rosswog S., Kasen D. 2008, arXiv:0811.1370v2 [astro-ph]
\bibitem[Haring \& Rix(2004)]{haringrix} Haring N., Rix H.W. 2004, \apj, 604, L89
\bibitem[Heckman et al.(2004)]{heckman} Heckman T.M., Kauffmann G., Brinchmann J., Charlot S., Tremonti C., White S.D.M. 2004, \apj, 613, 109 
\bibitem[Hirose et al.(2009)]{hirose} Hirose S., Krolik J., Blaes O.  2009, \apj, 691, 16
\bibitem[Hopkins et al.(2007)]{hopkins} Hopkins P.F., Richards, G.T., Hernquist, L. 2007, \apj, 654, 731
\bibitem[Khokhlov \& Melia(1996)]{khokhlov} Khokhlov A., Melia F. 1996, \apjl, 457, L61
\bibitem[Kochanek(1994)]{kochanek} Kochanek C.S. 1994, \apj, 422, 508
\bibitem[Komossa(2002)]{komossa} Komossa S. 2002, Rev. Mod. Astron., 15, 27
\bibitem[Koratkar \& Blaes(1999)]{kblaes} Koratkar A., Blaes O. 1999, PASP, 111, 1
\bibitem[Lacy et al.(1982)]{lacy} Lacy J.H., Townes C.H., Hollenbach D.J. 1982, \apj, 262, 120
\bibitem[Li et al.(2002)]{li} Li L., Narayan R., Menou K. 2002,
  \apj, 576, 753
\bibitem[Lodato et al.(2009)]{lodato} Lodato G., King A.R., Pringle J.E. 2009, \mnras, 392, 332
\bibitem[Loeb \& Ulmer(1997)]{loeb} Loeb A., Ulmer A. 1997, \apj, 489, 573
\bibitem[Magnier(2007)]{magnier} Magnier E. 2007, ASPC, 364, 153, ed. C. Sterken (San Francisco: ASP)
\bibitem[Magorrian \& Tremaine(1999)]{magtrem} Magorrian J., Tremaine S. 1999, \mnras, 309, 447
\bibitem[Merritt \& Poon(2004)]{mp} Merritt D., Poon M.Y. 2004, \apj, 606, 788
\bibitem[Ohsuga et al.(2005)]{ohsuga} Ohsuga K., Mori M., Nakamoto T., Mineshige S.  2005, \apj, 628, 368 
\bibitem[Perets et al.(2007)]{perets} Perets H.B., Hopman C., Alexander T. 2007, \apj, 656, 709
\bibitem[Phinney(1989)]{phinney} Phinney E.S. 1989, in IAU Symp. 136, The Galactic Center, ed. M. Morris (Dordrecht: Kluwer)
\bibitem[Ramirez-Ruiz \& Rosswog(2009)]{ramirez} Ramirez-Ruiz E., Rosswog S. 2009, \apjl, 697, L77
\bibitem[Rau et al.(2009)]{rau} Rau A., Kulkarni S.R., Law N.M. et al. 2009, arXiv:0906.5355v1 [astro-ph.CO]
\bibitem[Rauch \& Tremaine(1996)]{rauchtrem} Rauch K.P., Tremaine S. 1996, NewA, 1, 149
\bibitem[Rees(1988)]{rees} Rees, M. 1988, Nature, 333, 523  
\bibitem[Remillard \& McClintock(2006)]{remillard} Remillard R., McClintock J. 2006, \araa, 44
\bibitem[Rossi \& Begelman(2009)]{rossi} Rossi E.M., Begelman M.C.\ 2009, \mnras, 392, 1451 
\bibitem[Rybicki \& Lightman(1979)]{rybicki} Rybicki G.B., Lightman A.P.  1979, Radiative Processes in Astrophysics (New York:  John Wiley \& Sons, Inc.)
\bibitem[Shakura \& Sunyaev(1973)]{ss} Shakura N.I., Sunyaev  R.A. 1973, \aap, 24, 337
\bibitem[Ulmer(1999)]{ulmer99} Ulmer A. 1999, \apj, 514, 180
\bibitem[Zhao et al.(2002)]{zhao} Zhao H., Haehnelt M.G., Rees, M.J. 2002, NewA, 7, 385
\end{thebibliography}
\end{document}